\documentclass[12pt]{article}
\usepackage{epsfig,amsfonts,amssymb}
\usepackage{hyperref}
\usepackage{cite}
\input epsf.sty
\usepackage{cite}

\def\ZZZ{{\hbox{ Z\kern-1.6mm Z}}}
\def\RRR{{\hbox{ R\kern-2.4mm R}}}
\def\CCC{{\hbox{ C\kern-2.0mm C}}}
\def\zzz{{\hbox{z\kern-1mm z}}}

\newcommand{\qeq}{{\hbox{=\kern-2.3mm ? \kern.5mm }}}
\renewcommand{\qeq}{=}

\newcommand{\vp}{\varphi}

\newcommand{\GG}{{\cal G}}

\newcommand{\HH}{{\cal H}}
\newcommand{\MM}{{\cal M}}

\newcommand{\OO}{{\cal O}}

\newcommand{\PP}{{\cal P}}
\newcommand{\EE}{{\cal E}}

\newcommand{\wt}{\widetilde}
\newcommand{\wh}{\widehat}

\newcommand{\RR}{{\cal R}}

\newcommand{\be}{\begin{equation}}
\newcommand{\ee}{\end{equation}}
\newcommand{\ben}{\begin{eqnarray}\displaystyle}
\newcommand{\een}{\end{eqnarray}}

\newcommand{\refb}[1]{(\ref{#1})}
\newcommand{\p}{\partial}
\newcommand{\sectiono}[1]{\section{#1}\setcounter{equation}{0}}

\def\one{{\hbox{ 1\kern-.8mm l}}}
\def\zero{{\hbox{ 0\kern-1.5mm 0}}}

\newcommand{\bea}[1]{\begin{eqnarray}\label{#1} }
\newcommand{\eea}{\end{eqnarray}}

\newcommand{\eqref}{\refb}

\newcommand{\XX}{{\cal X}}

\usepackage{bm}
\usepackage[table]{xcolor}

\begin{document}

\baselineskip 24pt

\begin{center}
{\Large \bf  Gauge Invariant 1PI 
Effective Superstring Field Theory: Inclusion of the Ramond Sector}

\end{center}

\vskip .6cm
\medskip

\vspace*{4.0ex}

\baselineskip=18pt

\centerline{\large \rm Ashoke Sen}

\vspace*{4.0ex}

\centerline{\large \it Harish-Chandra Research Institute}
\centerline{\large \it  Chhatnag Road, Jhusi,
Allahabad 211019, India}

\vspace*{1.0ex}
\centerline{\small E-mail:  sen@mri.ernet.in}

\vspace*{5.0ex}

\centerline{\bf Abstract} \bigskip

We construct  
off-shell amplitudes in heterotic and type II 
string theories involving arbitrary combination
of Ramond and Neveu-Schwarz sector external states. We also construct the
equations of motion of 
a gauge invariant 1PI effective field theory which reproduces these off-shell
amplitudes. Using this construction we 
prove that the renormalized physical masses
do not
depend on the choice of local coordinate system and locations of picture changing
operators used in defining the off-shell amplitudes. We also use this formalism to
examine the conditions under which space-time
supersymmetry is unbroken in the quantum
theory.

\vfill \eject

\baselineskip=18pt

\tableofcontents

\sectiono{Introduction}

In a previous paper\cite{1411.7478}
we constructed a gauge invariant one particle irreducible (1PI) effective
action involving Neveu-Schwarz (NS)
sector fields of heterotic string theories and NSNS sector fields of type II string
theories. This led to a well defined algorithm for computing the renormalized 
masses of physical states. Furthermore it was possible to show that the
renormalized physical masses are independent of the choice of any spurious data
{\it e.g.} the choice of local coordinates at the punctures or the locations of the
picture changing operator (PCO)'s\cite{FMS} used in the construction of the 1PI action.
The goal of this paper is to extend the construction to the Ramond (R) sector 
for heterotic string theory and RNS, NSR and RR sectors of the type II 
string theory.

It has been known since the early days of string field theory\cite{wittenssft} that a straightforward
construction of a gauge invariant string field theory action involving R-sector states
is likely to fail due to the difficulty in the construction of the kinetic term
of the R-sector string fields. The difficulty has its origin in the fact that unlike in the
case of $-1$ picture NS sector states where the BPZ inner product between two
such states has the right picture number ($-2$) for giving a non-zero answer,  
the two
R sector states in the $-1/2$ picture  cannot have non-zero BPZ inner
product unless we insert additional operators of picture number $-1$ 
into the matrix element. This makes it
difficult to write down a kinetic term for the R sector fields that is local, 
commutes with $L_0^\pm$ so that it does not mix states at different levels and
whose cohomology coincides with the usual BRST cohomology.\footnote{It is possible to
live with this problem for open strings by working with kinetic operator which is
not diagonal in the $L_0$ basis as in \cite{wittenssft} 
but the problem reappears for closed string theory in which we are
interested.}
There is also an indirect
argument from low energy effective field theory which goes as follows. If we did
manage to write down a kinetic term for the R sector fields in a straightforward
manner then we could also use it to write down a gauge invariant kinetic term for
the RR 4-form field of type IIB string theory. But we know that it should not be possible to
write down a covariant action for the RR 4-form field due to the self-duality
constraint on its field strength.

We circumvent this problem by giving up the attempt to construct a gauge invariant
local 1PI action involving R-sector fields.
Instead we construct the gauge invariant 1PI effective
equations of motion.\footnote{A recent attempt  to construct the
equations of motion of R-sector fields in the Berkovits formulation of string field 
theory\cite{9503099,0109100,0406212,0409018} can be found in \cite{1312.7197,1412.5281}.} 
If we were trying to construct a string field theory action that
needs to be quantized then having equations of motion is not very useful -- one needs
the action for being able to quantize the theory. However the 1PI effective
theory by definition
already includes the effect of loop corrections and we are supposed to compute tree
amplitudes of this theory to find the full quantum corrected S-matrix of string 
theory. Thus
having the equations of motion of the 1PI effective theory is sufficient for our purpose.

We have not attempted to make the paper self-contained
-- it should be regarded as the completion of the program described in 
\cite{1408.0571,1411.7478}. Nevertheless
we review the main conventions in \S\ref{sback}. 
As the rest of the 
paper is mostly technical in nature, we shall try to summarize the main results here.
\begin{enumerate}
\item One of the bottlenecks faced in \cite{1408.0571,1411.7478} for 
generalizing the definition of off-shell
amplitudes to the R sector is finding a suitable definition of the gluing compatibility
condition. This is equivalent to the problem of finding a propagator in the R sector
if we restrict the R sector string fields to carry picture number $-1/2$. In \S\ref{soff}
we make a specific proposal where we insert into the usual NS sector propagator 
$b_0^+b_0^- (L_0^+)^{-1} \delta_{L_0^-}$ 
a factor of $\XX_0\equiv 
\ointop z^{-1} dz \XX(z)$ to define the propagator of R sector states in heterotic string 
theory.
Here
$\XX(z)$ is the picture changing operator and $b_0^\pm$ and $L_0^\pm$ are defined in
\refb{edefbpm}.  
The advantage of using the operator $\XX_0$ is that it commutes with 
$b_0^\pm$ and $L_0^\pm$\cite{9711087,1403.0940} and
hence can be inserted anywhere on the R sector propagator.
For the type II string we need a similar operator
$\bar\XX_0$ involving left handed PCO and insert $\XX_0$, $\bar\XX_0$ and $\XX_0\bar\XX_0$
into the propagator for $NSR$, $RNS$ and $RR$ fields. Once the propagators in 
different sectors are defined one can generalize the construction of off-shell amplitudes
in \cite{1408.0571,1411.7478}  to the 
Ramond sector states in a straightforward manner. This is discussed in
\S\ref{soff}.
\item We can use this definition of off-shell amplitudes to 
define 1PI amplitudes by restricting the integral over the moduli space to a restricted
domain such that the `1PI Riemann surfaces' associated with this restricted domain, 
together
with all other Riemann surfaces which can be obtained by plumbing fixture of the
1PI Riemann surfaces in all possible ways, generate all the Riemann surfaces over
which we integrate to get the full off-shell amplitude.
The generating functional of the 1PI amplitudes define the 1PI
effective field theory whose off-shell Green's functions in the Siegel gauge would agree
with the off-shell amplitudes constructed in \S\ref{soff}. This construction of 1PI effective
field theory is carried out in 
\S\ref{saction}. 
The general string field configuration $|\Psi\rangle$ is taken to be an element of the
Hilbert space of matter-ghost conformal field theory with picture number $-1$ in the 
NS sector and picture number $-1/2$ in the R sector.
The equations of motion are given in \refb{eeom} and its infinite
dimensional gauge invariance is described in \refb{egauge}. In these equations
$\GG$ stands respectively for the identity operator and $\XX_0$ 
while acting on the NS and R sector states of heterotic
string theory. The equations of motion and gauge transformation laws of type II string
theory have the same form with $\GG$ standing respectively
for the identity operator, $\XX_0$,
$\bar\XX_0$ and $\XX_0\bar\XX_0$ while acting on the NSNS, NSR, RNS and RR sector
states. 
\item Even though there is no fully satisfactory 1PI effective action for this theory,
in \S\ref{smatrix} we show that it is possible to write down
an action \refb{eact} from which we can derive the equations of motion. The problem with this
action is that 
{\it it contains extra  states that are not present in string theory.} 
For a classical theory that needs to be
quantized, the presence of these extra states would be fatal since they would propagate
in the loop and completely change the loop amplitudes. However since the
1PI theory is to be used only for classical / tree level computation, we can use this action to 
compute the S-matrix elements of string theory 
by restricting the external states to a subset of states which
correspond to genuine physical states in string theory.
\item The definition of $[~]$ used in \refb{eeom}, \refb{egauge} depends on the
choice of the local coordinate system at the punctures and the PCO locations used
in defining the off-shell amplitudes. In \S\ref{seffect} we show that the change in the equations
of motion \refb{eeom} under these changes can be absorbed into a redefinition of the
string field $|\Psi\rangle$. This is turn shows that the physical renormalized masses
are independent of the choices of local coordinate systems and PCO locations, generalizing 
the results of \cite{1311.1257,1401.7014}.
\item The gauge transformation laws \refb{egauge} automatically include local 
supersymmetry transformations. In \S\ref{ssusy} we discuss the conditions under 
which there is unbroken global supersymmetry. Our analysis leads to a condition similar
to the one found in \cite{1209.5461,1304.2832}, 
except that we arrive at a slightly different procedure for dealing with
divergences associated with separating type degenerations compared to the one
suggested in \cite{1209.5461,1304.2832}.
\item We conclude in \S\ref{sdiss} by discussing possible future applications of this
approach -- study of non-perturbative effects in string theory and the study of string
theory in RR background field.
\end{enumerate}

Finally, as in \cite{1411.7478}, we would like to emphasize that even though we have used
the PCO formalism for the construction of the off-shell amplitudes and 1PI effective theory,
it may also be possible to carry out similar construction in the more geometric approach
where superstring amplitudes are represented as integrals over supermoduli
spaces. For on-shell amplitudes such a formalism has already been developed (see 
\cite{Belopolsky,9706033,dp,1209.5461,1304.2832,Witten,donagi-witten,1403.5494}
for recent developments). It is also conceivable that once such a formalism is
developed for off-shell amplitudes and 1PI effective theory, one should also be able to
show its 
equivalence with the formalism developed here based on picture changing operators.

\sectiono{Conventions and definitions} \label{sback}

We shall follow the notations of \cite{1408.0571,1411.7478}. 
We begin our discussion with
heterotic string theory. In this case the
world sheet theory contains a
matter superconformal field theory with
central charge (26,15), and a ghost system of  total central charge $(-26,-15)$
containing anti-commuting $b$, $c$, $\bar b$, $\bar c$
ghosts and commuting $\beta, \gamma$ ghosts. Of these $b,c,\beta,\gamma$ are
right-handed and $\bar b,\bar c$ are left-handed.
The $(\beta,\gamma)$ system can be bosonized as\cite{FMS}
\be \label{eboserule}
\gamma = \eta\, e^{\phi}, \quad \beta= \p\xi \, e^{-\phi}, \quad \delta(\gamma)
= e^{-\phi}, \quad \delta(\beta) = e^\phi\, ,
\ee
where $\xi, \eta$ are  fermions and $\phi$ is a scalar with background charge.
The (ghost number, picture number, GSO) quantum numbers carried by various fields
are as follows:
\ben \label{equantum}
&& c, \bar c: (1,0,+), \quad b, \bar b: (-1, 0,+), \quad \gamma: (1,0,-), \quad \beta:(-1,0,-),
\nonumber \\
&& \xi: (-1,1,+), \quad \eta: (1,-1,+), \quad
e^{q\phi}: (0, q, (-1)^q)\, .
\een
We denote by $Q_B$ the BRST operator of this
theory and by $\XX(z)$ the picture changing operator
\be \label{epco}
\XX(z) = \{Q_B, \xi(z)\} = c \partial \xi + 
e^\phi T_F - {1\over 4} \p\eta e^{2\phi} b
- {1\over 4} \p\left(\eta e^{2\phi} b\right)\, .
\ee
This is a BRST invariant dimension zero primary operator 
and carries picture number $1$.

We now introduce vector spaces 
$\HH_{(n)}$ containing a subset of GSO even states in the matter-ghost
conformal field theory
satisfying the following conditions:
\be \label{econd}
|s\rangle \in \HH_{(n)} \quad \hbox{iff} \quad b_0^-|s\rangle = 0, \quad L_0^{-}|s\rangle =0\, ,
\quad \eta_0 |s\rangle =0, 
\quad \hbox{picture number of $|s\rangle=n$}\, ,
\ee
where
\be \label{edefbpm}
b_0^\pm \equiv(b_0\pm\bar b_0), \quad L_0^\pm\equiv(L_0\pm\bar L_0), \quad
c_0^\pm = {1\over 2} (c_0\pm\bar c_0)\, .
\ee
Note that $\HH_{(n)}$ contains NS-sector states for $n\in\ZZZ$ and R-sector states
for $n\in\ZZZ+{1\over 2}$. 
Although eventually we shall be interested in states for which the coefficient of the
NS-sector states are even elements of the grassmann algebra and the coefficients 
of the R-sector states are odd elements of the grassmann algebra, for now we shall
work with a more general space in which we allow the coefficients in each $\HH_{(n)}$
to be a general element of the grassmann algebra.
%Instead of taking them to be vector space over real fields, we shall regard the elements
%of $\HH_{(n)}$ for even $n$ to be linear combination of the basis states 
%of $\HH_{(n)}$ with grassmass even coefficients and the elements of
%$\HH_{(n)}$ for odd 
%$n$ to be linear combination of the basis states of $\HH_{(n)}$ 
%with grassmass odd coefficients.
We shall also define
\be\label{ecrude}
\HH_{NS}=\oplus_{n\in\zzz} \HH_{(n)}, \quad \HH_R = \oplus_{n\in\zzz+{1\over 2}} \HH_{(n)}
, \quad
\HH_T=\HH_{NS}\oplus \HH_R\, .
\ee
In the construction of string field theory a general off-shell string field configuration
will be represented by an
element of $\HH_{T}$ with ghost number 2 and 
picture numbers $-1$ or $-1/2$.
However for now we shall work with general states in $\HH_T$.

Next we introduce the operator -- introduced earlier 
in \cite{9711087,1312.2948,1403.0940,1407.8485}
for construction of string field theory action in the NS sector,
\be \label{edefggr}
\XX_0 = \ointop {dz\over z}\, \XX(z)
\ee
where the integration runs around an anti-clockwise contour enclosing the origin
with the factor of $1/2\pi i$ included in its definition. We need to treat $\XX_0$ as
an operator in radial quantization, acting on states represented by vertex operators at
the origin.  
The important properties of $\XX_0$ are its
commutation relations\cite{9711087,1403.0940}
\be\label{egridentity}
[b_0, \XX_0]=0, \quad [L_0, \XX_0]=0, \quad 
[\bar b_0, \XX_0]=0, \quad [\bar L_0, \XX_0]=0,  \quad [Q_B,\XX_0]=0\, .
\ee
The first identity requires some discussion. Using \refb{epco} we get
\be
[b_0, \XX_0] = \ointop dz \,\p\xi(z)\, .
\ee
This would vanish if $\xi(z)$ is single valued. Now even though $\xi(z)$ is not
an allowed conformal field in the small Hilbert space that we are working in\cite{FMS}
-- encoded in the $\eta_0|s\rangle=0$ condition in \refb{econd} -- it was shown in
\cite{Verlinde:1987sd} that all the correlation functions of $\xi(z)$ on 
arbitrary Riemann surfaces
are indeed single-valued. In terms of operators in the small Hilbert space this means
that $\ointop dz \p\xi(z)$ vanishes for integration over any closed contour on the
Riemann surface. This leads to the first equations in \refb{egridentity}. The other 
equations follow in a straightforward manner.

For convenience we shall define the general operator $\GG$ acting on
$\HH_T$ as
\be \label{edefgg}
\GG|s\rangle =\begin{cases} {|s\rangle \quad \hbox{if $|s\rangle\in \HH_{NS}$}\cr
\XX_0\, |s\rangle \quad \hbox{if $|s\rangle\in \HH_R$}}
\end{cases}\, .
\ee

For type II string theories we also have left-handed
commuting ghosts $\bar\beta,\bar\gamma$ which can be bosonized as in
\refb{eboserule}, introducing the fields $\bar\xi,\bar\eta,\bar\phi$. We also need to
introduce left-handed GSO quantum numbers and picture numbers and declare that
the right-handed fields are neutral under the left-handed GSO and left-handed
picture numbers while the left-handed fields are neutral under the right-handed GSO
and right-handed picture numbers. However for the ghost number we do not
distinguish between left and right handed sectors so that the 
$\bar\xi,\bar\eta,e^{q\bar\phi}$ carry the same ghost numbers as their right-handed
counterpart.
We introduce the left-handed PCO
\be \label{epcoii}
\bar\XX(\bar z) = \{Q_B, \bar\xi(\bar z)\} = \bar c \bar \partial \bar \xi + 
e^{\bar \phi} \bar T_F - {1\over 4} \bar \p\bar \eta e^{2\bar \phi} \bar b
- {1\over 4} \bar \p\left(\bar \eta e^{2\bar \phi} \bar b\right)\, ,
\ee
and 
\be
\bar\XX_0 = \ointop {d\bar z\over \bar z} \bar\XX(\bar z)\, .
\ee
The relevant states in the Hilbert space can now be divided into the subspaces
$\HH_{(m,n)}$ where $m$ and $n$ denote respectively the left and the
right-handed picture numbers. Each sector contains states that are annihilated by
$L_0^-$, $b_0^-$, $\eta_0$ and $\bar\eta_0$. 
The analog of \refb{ecrude} is
\ben\label{ecrudeii}
&& \HH_{NSNS}=\oplus_{m,n\in\zzz} \HH_{(m,n)}, \quad 
\HH_{NSR}=\oplus_{m\in\zzz,n\in\zzz+{1\over 2}} \HH_{(m,n)},\nonumber \\ &&
\HH_{RNS}=\oplus_{m\in\zzz+{1\over 2},n\in\zzz} \HH_{(m,n)},\quad 
\HH_{RR}=\oplus_{m,n\in\zzz+{1\over 2}} \HH_{(m,n)}, \nonumber \\ &&
\HH_T=\HH_{NSNS}\oplus \HH_{NSR}\oplus \HH_{RNS}\oplus \HH_{RR}\, .
\een
Finally we define 
\be \label{edefggii}
\GG|s\rangle =\begin{cases} {|s\rangle \quad \hbox{if $|s\rangle\in \HH_{NSNS}$}\cr
\XX_0\, |s\rangle \quad \hbox{if $|s\rangle\in \HH_{NSR}$}\cr 
\bar\XX_0\, |s\rangle \quad \hbox{if $|s\rangle\in \HH_{RNS}$}\cr 
\XX_0\bar\XX_0\, |s\rangle \quad \hbox{if $|s\rangle\in \HH_{RR}$}\cr 
}\, .
\end{cases}
\ee

\def\n{n}

\def\m{m}

\sectiono{Off-shell amplitudes}  \label{soff}

$g$-loop, $n$-point  on-shell amplitude in bosonic string theory is expressed as an
integral over the $(6g-6+2n)$ dimensional 
moduli space $\MM_{g,n}$ of genus $g$ Riemann surfaces with
$n$-punctures. Defining off-shell amplitudes in bosonic string theory requires extra
data in the form of a choice of local coordinate system around each puncture. This
requires us to introduce an infinite dimensional space $\wh\PP_{g,n}$ with the structure
of a fiber bundle whose base is $\MM_{g,n}$ and whose (infinite dimensional) 
fiber is parametrized by the possible choices of local
coordinate system around each puncture\cite{nelson,9206084}. 
The off-shell amplitude is described as an 
integral of a $(6g-6+2n)$-form over a section of $\wh\PP_{g,n}$. The construction of the
differential form to be integrated as well as the subspaces over which we need to
integrate can be found in \cite{1408.0571} (and also reviewed in \cite{1411.7478}).

Defining off-shell
amplitude in heterotic and type II string theories requires even more data -- a choice of the
locations of certain number of PCO's on the Riemann surface.
Let us for definiteness focus on the heterotic string theory -- generalization to type II string
theories will be discussed later.
A genus $g$ amplitude in heterotic string theory with $\m$ NS sector external states in the
$-1$ picture
and $\n$ Ramond sector external states in the $-1/2$ picture requires a total of 
$2g-2+\m + \n/2$ PCO insertions. Even though we shall need to relax the 
constraint on the picture number on the states for various manipulations, 
the off-shell amplitudes that we shall need will always involve 
NS-sector external 
states in the $-1$ picture and R-sector external states in the $-1/2$ picture, and hence
we shall always use 
the same number of PCO insertions on a genus $g$ Riemann surface with $m$ NS-sector
external states and $n$ R-sector external states. 
Thus we need to introduce a bigger fiber-bundle
$\wt\PP_{g,\m,\n}$ with $\MM_{g,\m,\n}$ -- the moduli space
of genus $g$ Riemann surface with $m$ NS-punctures and $n$ R-punctures -- 
as base and 
the choice of local coordinates at the
punctures and the $2g-2+\m + \n/2$  
PCO locations as fibers. The off-shell amplitude is defined as an 
integral of an appropriate differential form of degree $6g-6+2(m+n)$ 
over an appropriate $6g-2+2(m+n)$-dimensional subspace 
of this fiber bundle (together with a sum over spin structures which we shall
include in the definition of the integral). In fact,
following the procedure described in \cite{1408.0571} we can explicitly construct a set of 
$p$ forms $\Omega^{(g,\m,\n)}_p(|\tilde\phi_1\rangle,\cdots 
|\tilde\phi_{m}\rangle,|\hat\phi_1\rangle,\cdots |\hat\phi_n\rangle)$ 
on $\PP_{g,\m,\n}$ 
for all $p$, satisfying useful identities
to be discussed 
in \refb{eomega}, \refb{exchange}, \refb{efactor}, \refb{efactoraa},
whose the first $m$ arguments $|\tilde\phi_1\rangle,\cdots |\tilde\phi_m\rangle$ 
are NS sector states and the last $n$ arguments $|\hat\phi_1\rangle,\cdots |\hat\phi_n\rangle$ 
are R-sector states. 
$\Omega^{(g,\m,\n)}_p(|\tilde\phi_1\rangle,\cdots |\tilde\phi_m\rangle, |\hat\phi_1\rangle,\cdots
|\hat\phi_{\n}
\rangle)$ is expressed in terms of a correlation function in the matter-ghost CFT
with the state $|\tilde\phi_i\rangle\in\HH_{NS}$ and $|\hat\phi_j\rangle\in\HH_{R}$ 
inserted at the $i$-th and $(m+j)$-th punctures together with 
$p$ additional insertion of $b$ or $\bar b$ ghost fields and $2g-2+\m + \n/2$  
PCO insertions.
Ghost and picture number conservations tell us that $\Omega^{(g,\m,\n)}_p$ is
non-zero only if the total ghost number of $|\tilde\phi_1\rangle,\cdots |\tilde\phi_{\m}\rangle,
|\hat\phi_1\rangle,\cdots |\hat\phi_n\rangle$
is $p-6g+6$ and the total picture number of 
$|\tilde\phi_1\rangle,\cdots |\tilde\phi_{\m}\rangle,
|\hat\phi_1\rangle,\cdots |\hat\phi_n\rangle$
is $-\m-{1\over 2}\n$.\footnote{Even though
the emphasis in \cite{1408.0571} 
was on the NS sector external states for reasons to be explained
below, the construction of $\Omega^{(g,\m,\n)}_p(|\tilde\phi_1\rangle,\cdots |\tilde\phi_m\rangle,
|\hat\phi_1\rangle, \cdots
|\hat\phi_{\n}
\rangle)$ itself can be carried out in an identical manner irrespective of whether the
external states are NS or R-sector states.}

Compared to the case
of bosonic string theory, there are some new subtleties that arise in the choice of the subspace
of $\wt\PP_{g,\m,\n}$ over which we integrate. These are listed below:
\begin{enumerate}
\item As discussed in \cite{1408.0571}, generically the subspace of $\wt\PP_{g,\m,\n}$ over which
we need to integrate contains vertical segments -- along which the location of the PCO's
change at fixed values of the coordinates of the base $\MM_{g,\m,\n}$ --
in order to avoid spurious singularities\cite{Verlinde:1987sd,lechtenfeld,morozov}. 
The procedure for carrying out integrals over
these vertical segments was described in \cite{1408.0571} 
and works equally well for   NS or
R sector external states. Since subspaces containing vertical segments are not
strictly sections of $\wt\PP_{g,\m,\n}$, we refer to these as integration cycles.
\item It may not always be possible
to have a fixed subspace of $\wt\PP_{g,\m,\n}$ that is consistent with all the symmetries 
{\it e.g.} modular invariance
and symmetry under the permutations of external punctures.\footnote{Here we shall only
demand symmetry under the permutation of the NS sector punctures and separately
under the permutation of R-sector punctures.}
In order to deal with this
problem we allow the integration cycle to be formal weighted average of several 
subspaces. The
integral of a form on a formal weighted average of several subspaces is defined as the
weighted average of the integrals of the form over different subspaces.
From now on, when we refer to subspaces of $\wt\PP_{g,\m,\n}$, they will in general 
mean weighted average of subspaces.
\item \label{p3} 
The third subtlety arises while dealing with off-shell amplitudes with Ramond sector
external states. The problem has its origin in the fact that in order to ensure that the 
off-shell amplitude leads to sensible definition of physical quantities -- 
{\it e.g.} renormalized
physical masses and S-matrix elements --  
we need to ensure that the choice of the integration cycle is gluing compatible. To see
what it means, recall 
that if we consider two Riemann surfaces $\Sigma_1$ and $\Sigma_2$, and pick one
puncture on each of them with local coordinates $z$ and $w$, then we can  construct a two
parameter 
family of Riemann surfaces $\Sigma$
by joining $\Sigma_1$ and $\Sigma_2$ using the plumbing fixture 
relation:
\be \label{eplumb}
z\, w = e^{-s+i\theta}\, , \qquad 0\le s<\infty, \quad 0\le\theta<2\pi\, .
\ee
Gluing compatibility requires that on the resulting Riemann surfaces $\Sigma$ the 
choice of local coordinates as well as the locations of the PCO's will be induced by
those from the original Riemann surfaces $\Sigma_1$ and $\Sigma_2$. This does not
cause any problem when the two punctures which are glued are associated with 
NS-sector states. To see this we note that if $\Sigma_1$ has genus $g_1$ with $n_{N1}$
NS punctures and $n_{R1}$ R punctures and  $\Sigma_2$ has genus $g_2$ with $n_{N2}$
NS punctures and $n_{R2}$ R punctures, then we have $2g_1-2+n_{N1}+n_{R1/2}$
PCO's on $\Sigma_1$ and $2g_2-2+n_{N2}+n_{R2/2}$
PCO's on $\Sigma_2$. The sum of these matches the required number of PCO's on
$\Sigma$ which has genus $g_1+g_2$, $(n_{N_1}+n_{N2}-2)$ NS-punctures and
$(n_{R1}+n_{R2})$ R-punctures. However if the two punctures being glued are 
of R-type, then $\Sigma$ has genus $g_1+g_2$, $(n_{N_1}+n_{N2})$ NS-punctures and
$(n_{R1}+n_{R2}-2)$ R-punctures. The required number of PCO's on $\Sigma$
is $2(g_1+g_2) + 2(n_{N1}+n_{N2}) + (n_{R1}+n_{R2})/2 -3$ which is one more than
the total number of PCO's on $\Sigma_1$ and $\Sigma_2$. For this reason the analysis
in \cite{1408.0571} was restricted mostly to NS sector external states.
\end{enumerate}

In this paper we propose a prescription for the choice of the PCO's on $\Sigma$ when the
punctures being glued are Ramond punctures. Our prescription will be to choose
$2(g_1+g_2) + 2(n_{N1}+n_{N2}) + (n_{R1}+n_{R2})/2 -4$ of the PCO locations to be
those induced from $\Sigma_1$ and $\Sigma_2$ and the last PCO to be $\XX_0$ given
in \refb{edefggr}. In other words we do not insert the extra PCO at a single point but take
a formal weighted average of infinite number of insertions given by
\be \label{eaverage}
\XX_0\equiv \ointop {dz\over z} \XX(z)\, .
\ee
The contour of integration can be taken to be any anti-clockwise contour with 
$e^{-s}\le |z|\le 1$. This translates to the same condition on $w=e^{-s+i\theta}/z$. 
Furthermore even though
$dz/z=-dw/w$, an anti-clockwise contour in the $z$-plane corresponds to a clockwise
contour in the $w$ plane. Thus the prescription is symmetric between the two punctures.

Once a gluing compatible integration cycle has been chosen this way, we can define the 
off-shell amplitude for $m$ NS-sector external states $|\tilde\phi_1\rangle,\cdots |\tilde\phi_m\rangle$
and $n$ R-sector external states $|\hat\phi_1\rangle,\cdots |\hat\phi_n\rangle$ by integrating
$\Omega^{(g,m,n)}_{6g-6 + 2(m+n)}(|\tilde\phi_1\rangle,\cdots |\tilde\phi_m\rangle,
|\hat\phi_1\rangle,\cdots |\hat\phi_n\rangle)$ over this integration cycle.
Of course in order to prove the usefulness of this prescription we need to show that the
physical quantities computed from this prescription, {\it e.g.} the renormalized masses
and S-matrix elements, are independent of the choice of the integration cycles used in defining
the off-shell amplitude. This will be done in the next sections by 
turning this into a prescription for constructing a gauge invariant effective field theory and then
showing that the effect of changing the integration cycles can be absorbed into a field
redefinition.

We now address a few issues associated with this prescription:
\begin{enumerate}
\item
Suppose we have chosen the locations of the PCO's on $\Sigma_1$ and $\Sigma_2$
so as to avoid spurious poles. Is it guaranteed that the 
relevant correlation function on $\Sigma$, with the PCO arrangements as described
above, is free from spurious singularities? As described in \cite{1408.0571}, if we 
choose the local coordinates $z$ and $w$ at the punctures being glued in such a way
that $|z|\le 1$ and $|w|\le 1$ describe sufficiently small disks around the respective
punctures, then the relevant correlation function on $\Sigma$ is given approximately
by the sum of products of correlation functions on $\Sigma_1$ and $\Sigma_2$ 
with states of low $L_0^+$ inserted at the punctures that are being glued and the
matrix element of $b_0^+ b_0^- (L_0^+)^{-1}$ (or 
$b_0^+ b_0^- (L_0^+)^{-1} \XX_0$) between these low $L_0^+$ states 
if they belong to the NS sector (or R-sector). 
This is free from spurious singularity by construction. We shall always choose the
local coordinate systems at the punctures in this manner. 
\item The
family of Riemann surfaces 
described in \refb{eplumb} has a boundary at $s=0$. This is not a boundary of the moduli space
and hence the full
integration cycle must involve Riemann surfaces which lie beyond this boundary.
On these Riemann surfaces the choice of PCO's is not restricted by the choice of locations
of the PCO's on $\Sigma_1$ and $\Sigma_2$ except that we require the choice of PCO
locations to be continuous across this boundary. If we require the choice of PCO locations
to be continuous everywhere in the moduli space then we'll need to continue using the 
averages over PCO locations like the one given in \refb{eaverage} everywhere in the
moduli space. However this is not necessary, since using the rules for 
`integration across vertical
segments' we can allow the PCO locations to jump discontinuously across
codimension one subspaces of the moduli space. We now give an example of such a
construction. With the choice of local
coordinates of the type described above we expect that we can continue to choose the
PCO insertions of the type we have used till $s=-\epsilon$ for sufficiently small $\epsilon$
without encountering any spurious pole.
Now we can choose the PCO locations such that over the codimension one subspace of the
moduli space given by $s=-\epsilon$
all the PCO locations labelled by $z$ in \refb{eaverage}
change to some fixed value $z_0$. According to the
prescription given in \cite{1408.0571} this will 
require integrating over this subspace of the moduli
space an appropriate differential form whose construction involves the insertion of
(see {\it e.g.} eq.(3.45) of \cite{1408.0571})
\be
\ointop {dz\over z} (\xi(z) - \xi(z_0))
\ee
into the correlation function. In this case beyond the $s=-\epsilon$ subspace we
can use the extra PCO location to be at some fixed point $z_0$ instead of being 
distributed over a circle. More generally the integration cycle can contain different
segments in which the character of the PCO locations could change, with some 
segments containing all the PCO locations at fixed points on the Riemann surface, 
while the other segments having
one or more of the PCO locations averaged over insertions over one (or even two)
 dimensional
subspaces of the Riemann surface.
\end{enumerate}

We can now proceed in a manner identical to that
in \cite{1408.0571,1411.7478} and introduce the  $6g-6+2(m+n)$ dimensional subspaces
$\RR_{g,\m,\n}$ of $\wt\PP_{g,\m,\n}$, known as 1PI subspaces, such that by gluing
the Riemann surfaces associated with $\RR_{g,\m,\n}$ in all possible ways using the
plumbing fixture relation \refb{eplumb} we generate all Riemann surfaces associated
with the full integration cycle used to define the off-shell amplitudes.
We list below the important properties of $\Omega^{(g,\m,\n)}_p$ and
$\RR_{g,m,n}$. 

We begin with the properties of $\Omega^{(g,\m,\n)}_p$. First of all, we have
\ben \label{eomega}
&& \sum_{i=1}^{m} (-1)^{\tilde\gamma_1+\cdots \tilde\gamma_{i-1}} 
\Omega^{(g,m,n)}_p(|\tilde\phi_1\rangle, \cdots 
|\tilde\phi_{i-1}\rangle,
Q_B|\tilde\phi_i\rangle, |\tilde\phi_{i+1}\rangle, \cdots |\tilde\phi_{m}\rangle, |\hat\phi_1\rangle,\cdots |\hat\phi_n\rangle)
\nonumber \\
&& + \sum_{i=1}^{n} (-1)^{\tilde\gamma_1+\cdots \tilde\gamma_{m}+\hat \gamma_1+
\cdots \hat \gamma_{i-1}} 
\Omega^{(g,m,n)}_p(|\tilde\phi_1\rangle, \cdots |\tilde\phi_m\rangle, |\hat\phi_1\rangle, \cdots 
|\hat\phi_{i-1}\rangle,
Q_B|\hat\phi_i\rangle, |\hat\phi_{i+1}\rangle, \cdots |\hat\phi_{n}\rangle)
\nonumber \\
&=& (-1)^p d\Omega^{(g,m,n)}_{p-1}
(|\tilde\phi_1\rangle,\cdots, |\tilde\phi_{m}\rangle, |\hat\phi_1\rangle, \cdots |\hat\phi_n\rangle) \, ,
\een
where $d$ denotes exterior derivative on $\wt\PP_{g,\m,\n}$ and
\be
\tilde\gamma_i = \hbox{grassmannality of $|\tilde\phi_i\rangle$}, \quad
\hat \gamma_i = \hbox{grassmannality of $|\hat\phi_i\rangle$} \, ,
\ee
the grassmannality of an operator being defined as 0 (1) if the operator is
grassmann even (odd).
The grassmannality of a GSO even operator in the matter ghost
conformal field theory is equal to its ghost number mod 2 in the NS sector and
ghost number+1 mod 2 in the R sector if the coefficient multiplying the operator
is grassmann even. If the coefficient is grassmann odd then the grassmannality
will be opposite.
In the same convention,
$\Omega^{(g,\m,\n)}_p$ has the symmetry property
\ben \label{exchange}
&& s_{i,i+1}\circ \Omega^{(g,\m,\n)}_p(|\tilde\phi_1\rangle, \cdots |\tilde\phi_{i-1}\rangle,
|\tilde\phi_{i+1}\rangle, |\tilde\phi_{i}\rangle,|\tilde\phi_{i+2}\rangle\cdots  |\tilde\phi_{\m}\rangle,
 |\hat\phi_1\rangle, \cdots |\hat\phi_n\rangle
)\nonumber \\
 &=& (-1)^{\tilde\gamma_i \tilde\gamma_{i+1}} \, \Omega^{(g,\m,\n)}_p(|\tilde\phi_1\rangle, \cdots |\tilde\phi_m\rangle,
|\hat\phi_1\rangle,\cdots |\hat\phi_n\rangle)\nonumber \\
&& s_{m+i,m+i+1}\circ \Omega^{(g,\m,\n)}_p(|\tilde\phi_1\rangle, \cdots  |\tilde\phi_m\rangle,
|\hat\phi_1\rangle, \cdots |\hat\phi_{i-1}\rangle,
|\hat\phi_{i+1}\rangle, |\hat\phi_{i}\rangle,|\hat\phi_{i+2}\rangle\cdots  |\hat\phi_{\n}\rangle)
\nonumber \\
 &=& (-1)^{\hat\gamma_i \hat\gamma_{i+1}} \, 
 \Omega^{(g,\m,\n)}_p(|\tilde\phi_1\rangle, \cdots |\tilde\phi_m\rangle,
|\hat\phi_1\rangle,\cdots |\hat\phi_n\rangle)\, ,
\een
where $s_{i,i+1}$ is the transformation on $\wt\PP_{g,\m,\n}$ that exchanges the punctures
$i$ and $i+1$ together with their local coordinates
and $s_{i,i+1}\circ \Omega^{(g,\m,\n)}_p$ is the pullback of
$\Omega^{(g,\m,\n)}_p$ under this transformation.

Let us now turn to the properties of $\RR_{g,\m,\n}$.  First of all, $\RR_{g,m,n}$ is
taken to be
symmetric under the exchange of any pair of NS-punctures and also under the
exchange of any pair of R-punctures. This needs to be achieved, if necessary, by
taking $\RR_{g,m,n}$ to be formal weighted average of subspaces related by these
exchange transformations. 
Plumbing fixture of $\RR_{g_1,\m_1,\n_1}$ and
$\RR_{g_2,\m_2,\n_2}$ at an NS puncture produces a subspace of 
$\wt\PP_{g_1+g_2,\m_1+\m_2-2,
\n_1+\n_2}$ which we shall denote by $\RR_{g_1,\m_1,\n_1}\circ \RR_{g_2,\m_2,\n_2}$. 
On the other hand
plumbing fixture of $\RR_{g_1,\m_1,\n_1}$ and
$\RR_{g_2,\m_2,\n_2}$ at an R puncture produces a subspace of 
$\wt\PP_{g_1+g_2,\m_1+\m_2,
\n_1+\n_2-2}$ which we shall denote by $\RR_{g_1,\m_1,\n_1}\star \RR_{g_2,\m_2,\n_2}$. 
Note that the insertion of the extra
PCO \refb{eaverage} is included in the definition of 
$\RR_{g_1,\m_1,\n_1}\star \RR_{g_2,\m_2,\n_2}$. 
For definiteness let us choose the convention that the plumbing fixture will always
be done with the last (NS or R) puncture of the first Riemann surface and the first
(NS or R) puncture of the second Riemann surface. 
Furthermore on the Riemann surfaces associated with 
$\RR_{g_1,\m_1,\n_1}\circ \RR_{g_2,\m_2,\n_2}$ the first set of $m_1-1$
NS-punctures and $n_1$ R-punctures will represent the punctures on the
surfaces corresponding to $\RR_{g_1,m_1,n_1}$ and the last set of
$m_2-1$ NS-punctures  and $n_2$ R-punctures will represent the punctures on the
surfaces corresponding to $\RR_{g_2,m_2,n_2}$.
A similar convention will be followed for the punctures on the surfaces associated
with $\RR_{g_1,\m_1,\n_1}\star \RR_{g_2,\m_2,\n_2}$.
The subspaces $\RR_{g_1,\m_1,\n_1}\circ \RR_{g_2,\m_2,\n_2}$ and
$\RR_{g_1,\m_1,\n_1}\star \RR_{g_2,\m_2,\n_2}$ have natural boundaries 
containing the Riemann surfaces obtained by setting $s=0$ in the plumbing
fixture relations \refb{eplumb}. 
We shall denote them  by $\{\RR_{g_1,\m_1,\n_1}, \RR_{g_2,\m_2,\n_2}\}$
and $\{\RR_{g_1,\m_1,\n_1}; \RR_{g_2,\m_2,\n_2}\}$ respectively.
Thus
$\{\RR_{g_1,\m_1,\n_1} , \RR_{g_2,\m_2,\n_2}\}$ represents the set of punctured Riemann
surfaces equipped with choice of local coordinates at the punctures and PCO
locations that we obtain by gluing the families of Riemann surfaces  corresponding
to $\RR_{g_1.\m_1,\n_1}$ and $\RR_{g_2,\m_2,\n_2}$ at NS punctures
using plumbing fixture relation \refb{eplumb}
with the parameter
$s$ set to zero. 
$\{\RR_{g_1,\m_1,\n_1} ;\RR_{g_2,\m_2,\n_2}\}$ has a similar interpretation except that
the plumbing fixture is done at Ramond punctures, and we insert an extra PCO
given by \refb{eaverage}  around the punctures.
The orientations of $\{A,B\}$ and $\{A;B\}$
will be defined by taking its volume form to be
$d\theta\wedge dV_A\wedge dV_B$ where $dV_A$ and $dV_B$ are volume forms on 
$A$ and $B$ respectively.

The boundaries of $\RR_{g,\m,\n}$ are of special interest. 
Since $\MM_{g,m,n}$ has boundaries
corresponding to separating and non-separating type degenerations, 
the fibers over these boundaries correspond to boundaries of $\wt\PP_{g,m,n}$.
If $\RR_{g,m,n}$
intersects these boundaries of $\wt\PP_{g,m,n}$ then these will form 
boundaries of $\RR_{g,m,n}$. But by construction $\RR_{g,m,n}$ does not intersect the
boundaries of $\wt\PP_{g,m,n}$ corresponding to separating type degenerations -- they all arise
from the $s\to\infty$ limit of the plumbing fixture of two or more 1PI Riemann surfaces and hence 
lie in the 1PR region of the full integration cycle.
On the other hand although $\RR_{g,m,n}$ does intersect the boundaries of
$\wt\PP_{g,m,n}$ corresponding to non-separating type degenerations, we shall 
ignore them since
integrals of total derivatives do not receive any boundary contribution from 
there\cite{berera,1307.5124}.
The other boundaries of $\RR_{g,m,n}$ lie in the interior of $\wt\PP_{g,m,n}$ 
and match the
$s=0$ boundary of the subspaces $\RR_{g_1,\m_1,\n_1}\circ \RR_{g_2,\m_2,\n_2}$
or $\RR_{g_1,\m_1,\n_1}\star \RR_{g_2,\m_2,\n_2}$ for appropriate choices of
$g_i,\m_i,\n_i$. 
This gives
\ben  \label{eboundary}
\p \RR_{g,\m,\n} &=& -{1\over 2} \sum_{g_1,g_2\atop g_1+g_2=g} 
\sum_{\m_1,\m_2\atop \m_1+\m_2 = m+2}
\sum_{\n_1,\n_2\atop \n_1+\n_2 = n}
{\bf S}[\{\RR_{g_1,\m_1,\n_1} , \RR_{g_2,\m_2,\n_2}\}] \nonumber \\ &&
-{1\over 2} \sum_{g_1,g_2\atop g_1+g_2=g} 
\sum_{\m_1,\m_2\atop \m_1+\m_2 = m}
\sum_{\n_1,\n_2\atop \n_1+\n_2 = n+2}
{\bf S}[\{\RR_{g_1,\m_1,\n_1} ; \RR_{g_2,\m_2,\n_2}\}]
\, ,
\een
where ${\bf S}$ denotes the
operation of summing over inequivalent permutations of external NS-sector punctures and
also external R-sector punctures. Thus for example 
${\bf S}[\RR_{g_1,\m_1,\n_1}\circ \RR_{g_2,\m_2,\n_2}]$ involves sum over
${m_1+m_2-2\choose m_1-1}$ inequivalent permutation of the external NS-sector
punctures and ${\n_1+\n_2\choose \n_1}$ inequivalent permutation of the external R-sector
punctures. The minus sign on the right hand side reflects that $\RR_{g,m,n}$,
$\RR_{g_1,\m_1,\n_1}\circ \RR_{g_2,\m_2,\n_2}$ and
$\RR_{g_1,\m_1,\n_1}\star \RR_{g_2,\m_2,\n_2}$ will all have to fit together so they they
form a subspace of the full integration cycle used for defining the off-shell amplitude.
Thus the boundary of $\RR_{g,m,n}$ will be oppositely oriented to those of 
$\RR_{g_1,\m_1,\n_1}\circ \RR_{g_2,\m_2,\n_2}$ and
$\RR_{g_1,\m_1,\n_1}\star \RR_{g_2,\m_2,\n_2}$.
The factors of $1/2$ account for the double counting due to the symmetry that exchanges
the two Riemann surfaces corresponding to $\RR_{g_1,m_1,n_1}$ and
$\RR_{g_2,m_2,n_2}$.

Following analysis similar to that in \cite{1408.0571,1411.7478} one
can show that on 
$\{\RR_{g_1,\m_1,\n_1} , \RR_{g_2,\m_2,\n_2}\}$, $\Omega^{(g_1+g_2,\m_1+\m_2-2,
\n_1+\n_2)}_p$
satisfies the factorization property
\ben \label{efactor}
&&\int_\theta \, \Omega^{(g_1+g_2,\m_1+\m_2-2,\n_1+\n_2)}_p(|\tilde\phi_1\rangle, \cdots 
|\tilde\phi_{m_1+m_2-2}\rangle, |\hat\phi_1\rangle, \cdots |\hat\phi_{n_1+n_2}\rangle)
\nonumber \\
&=& \sum_{p_1,p_2\atop p_1+p_2 = p-1} \tilde\sigma_1 \, \tilde\sigma_2\, \tilde\sigma_3\, \tilde\sigma_4\, \tilde\sigma_5\,
\, \Omega^{(g_1,\m_1,\n_1)}_{p_1}(|\tilde\phi_1\rangle,\cdots |\tilde\phi_{\m_1-1}\rangle, |\tilde\vp_r\rangle,
|\hat\phi_{1}\rangle \cdots |\hat\phi_{\n_1}
\rangle)
\nonumber \\
&& \qquad \qquad \qquad
\wedge\,  \, \Omega^{(g_2,\m_2,\n_2)}_{p_2}( |\tilde\vp^r\rangle, |\tilde\phi_{\m_1}\rangle,\cdots |\tilde\phi_{\m_1+\m_2-2}\rangle, |\hat\phi_{n_1+1}\rangle, \cdots |\hat\phi_{n_1+n_2}\rangle)
\een
where $\int_\theta$ denotes integration over the angular coordinate
$\theta$ appearing in the plumbing fixture relation \refb{eplumb}
and $\{|\tilde\vp_r\rangle\}$ and
$\{|\tilde\vp^r\rangle\}$ are a set of dual basis of $\HH_{NS}$ satisfying
\be \label{ebasis}
\langle \tilde\vp^r | c_0^-|\tilde\vp_s\rangle = \delta^r{}_s \qquad \Leftrightarrow \qquad
\langle \tilde\vp_s | c_0^-|\tilde\vp^r\rangle = \delta^r{}_s  \, .
\ee
$\tilde\sigma_1$ is a sign
that arises in changing the ordering of the vertex operators for\break \noindent
$|\tilde\phi_{m_1}\rangle, \cdots 
|\tilde\phi_{m_1+m_2-2}\rangle, |\hat\phi_1\rangle, \cdots |\hat\phi_{n_1}\rangle 
$ to 
$|\hat\phi_1\rangle, \cdots |\hat\phi_{n_1}\rangle,|\tilde\phi_{m_1}\rangle, \cdots 
|\tilde\phi_{m_1+m_2-2}\rangle$.
$\tilde\sigma_2$ is a sign factor that arises
in moving the vertex operator for $|\tilde\vp_r\rangle$ through those of
$|\hat\phi_1\rangle, \cdots |\hat\phi_{n_1}\rangle$. The rest of the sign factors
$\tilde\sigma_3,\tilde\sigma_4$ and $\tilde\sigma_5$ given in
\refb{efactor} were already present in \cite{1408.0571,1411.7478}
and originate from three sources. $\tilde\sigma_3$ arises because
we need to move $p_2$ of the
$b$-ghost insertions associated with 
$\Omega^{(g_2,\m_2,\n_2)}_{p_2}$
through the vertex operators of $
|\tilde\phi_1\rangle,\cdots |\tilde\phi_{\m_1-1}\rangle, 
|\hat\phi_{1}\rangle \cdots |\hat\phi_{\n_1}\rangle$. $\tilde\sigma_4$ arises from
the need to move the vertex operator of
$|\tilde\vp^r\rangle$ through the $p_2$ insertions of $b$ ghosts associated 
with $\Omega^{(g_2,\m_2,\n_2)}_{p_2}$. Finally $\tilde\sigma_5$ arises due to the need
to move a factor
of $b_0^-$ through the $p_1$ insertions of $b$-ghost operators associated with
$\Omega^{(g_1,\m_1,\n_1)}_{p_1}$ and the vertex operators of
$|\tilde\phi_1\rangle,\cdots |\tilde\phi_{\m_1-1}\rangle,
|\hat\phi_{1}\rangle \cdots |\hat\phi_{\n_1}\rangle$.

On the other hand
on $\{\RR_{g_1,\m_1,\n_1} ; \RR_{g_2,\m_2,\n_2}\}$, $\Omega^{(g_1+g_2,\m_1+\m_2,
\n_1+\n_2-2)}_p$
satisfies the factorization property
\ben \label{efactoraa}
&& \int_\theta \, \Omega^{(g_1+g_2,\m_1+\m_2,\n_1+\n_2-2)}_p(|\tilde\phi_1\rangle, \cdots 
|\tilde\phi_{m_1+m_2}\rangle, |\hat\phi_1\rangle, \cdots |\hat\phi_{n_1+n_2-2}\rangle)
\nonumber \\
&=& \sum_{p_1,p_2\atop p_1+p_2 = p-1} 
\hat\sigma_1 \, \hat\sigma_2\, \hat\sigma_3\, \hat\sigma_4\, \hat\sigma_5
\, \Omega^{(g_1,\m_1,\n_1)}_{p_1}(|\tilde\phi_1\rangle,\cdots |\tilde\phi_{\m_1}\rangle, 
|\hat\phi_{1}\rangle, \cdots |\hat\phi_{\n_1-1}
\rangle,
|\hat\vp_r\rangle)
\nonumber \\
&& \qquad \qquad 
\wedge\,  \, \Omega^{(g_2,\m_2,\n_2)}_{p_2}(
 |\tilde\phi_{\m_1+1}\rangle,\cdots |\tilde\phi_{\m_1+\m_2}\rangle, 
\XX_0 |\hat\vp^r\rangle,
 |\hat\phi_{n_1}\rangle, \cdots |\hat\phi_{n_1+n_2-2}\rangle) \nonumber \\
\een
where  $\{|\hat\vp_r\rangle\}$ and
$\{|\hat\vp^r\rangle\}$ are a set of dual basis of $\HH_{R}$
satisfying
\be \label{ebasisaa}
\langle \hat\vp^r | c_0^-|\hat\vp_s\rangle = \delta^r{}_s \qquad \Leftrightarrow \qquad
\langle \hat\vp_s | c_0^-|\hat\vp^r\rangle = \delta^r{}_s  \, .
\ee
$\hat\sigma_1$ denotes the sign picked up while moving the vertex operators of
$|\tilde\phi_{\m_1+1}\rangle,\cdots |\tilde\phi_{\m_1+\m_2}\rangle$ through those of
$|\hat\phi_{1}\rangle, \cdots |\hat\phi_{\n_1-1}\rangle$ and $\hat\sigma_2$ is the sign picked up while
moving the vertex operator for $\XX_0 |\hat\vp^r\rangle$ through those of 
$ |\tilde\phi_{\m_1+1}\rangle,\cdots |\tilde\phi_{\m_1+\m_2}\rangle$. 
$\hat\sigma_3$ arises because
we need to move $p_2$ of the
$b$-ghost insertions associated with 
$\Omega^{(g_2,\m_2,\n_2)}_{p_2}$
through the vertex operators of $
|\tilde\phi_1\rangle,\cdots |\tilde\phi_{\m_1}\rangle, 
|\hat\phi_{1}\rangle \cdots |\hat\phi_{\n_1-1}\rangle$. $\hat\sigma_4$ arises from
the need to move the vertex operator of
$\XX_0|\hat\vp^r\rangle$ through the $p_2$ insertions of $b$ ghosts associated 
with $\Omega^{(g_2,\m_2,\n_2)}_{p_2}$. Finally $\hat\sigma_5$ arises due to the need
to move a factor
of $b_0^-$ through the $p_1$ insertions of $b$-ghost operators associated with
$\Omega^{(g_1,\m_1,\n_1)}_{p_1}$ and the vertex operators of
$|\tilde\phi_1\rangle,\cdots |\tilde\phi_{\m_1}\rangle,
|\hat\phi_{1}\rangle \cdots |\hat\phi_{\n_1-1}\rangle$.

The derivation of \refb{efactoraa} follows in the same way as its NS-sector counterpart
\refb{efactor} described in \cite{1408.0571,1411.7478}. 
The extra factor of $\XX_0$ has its origin in the extra insertion of the PCO
given in \refb{eaverage} involving plumbing fixture of Ramond punctures.
Note that  using \refb{edefgg} 
we could replace $\XX_0$ by $\GG$ in \refb{efactoraa} and inserted a factor
of $\GG$ in front of $|\tilde\vp^r\rangle$ in \refb{efactor} to make the two equations look
similar. This will be exploited later.

Generalization to type II string theories 
requires effectively `doubling' the number of PCO's by including appropriate number
of PCO's from the left-handed sector. Now for degenerations at NSR, RNS and RR 
punctures we 
insert respectively extra factor of $\XX_0$, $\bar\XX_0$ and $\XX_0\bar\XX_0$ around the
punctures. Rest of the analysis proceeds in a straightforward manner.

\sectiono{The 1PI effective field theory} \label{saction}

We shall now construct the gauge invariant equations of motion of a 1PI
effective field theory whose off-shell amplitudes coincide with the ones constructed
in \S\ref{soff}. Again for simplicity we first focus on the heterotic string theory.
We shall begin by defining certain multilinear functions of the elements of $\HH_T$
motivated by related construction in
bosonic string field theory\cite{9206084}.

\subsection{The $\{~\}$ and $[~]$ products} 

We  define, for $|\Phi_i\rangle\in\HH_T$, a function $\{\Phi_1\cdots \Phi_N\}$
as follows.\footnote{We only need the definitions of $\{\Phi_1\cdots \Phi_N\}$ and
$[\Phi_1\cdots \Phi_N]$ in cases where each of the $|\Phi_i\rangle$'s belong to
$\HH_{-1}\oplus\HH_{-1/2}$ in heterotic string theory and $\HH_{(-1,-1)}\oplus
\HH_{(-1/2,-1)}\oplus \HH_{(-1,-1/2)}\oplus \HH_{(-1/2,-1/2)}$ in type II string theories.}
\begin{enumerate}
\item $\{\Phi_1\cdots \Phi_N\}$ is a multilinear function of $|\Phi_1\rangle,\cdots |\Phi_N\rangle$
taking values in the grassmann algebra.
Since we can express each $|\Phi_i\rangle$ as a linear combination of states in $\HH_{(n)}$,
it is enough to define $\{\Phi_1\cdots \Phi_N\}$ in the case where each $|\Phi_i\rangle$ is
either an NS sector state or an R-sector state and has a fixed grassmannality.
\item $\{\Phi_1\cdots \Phi_N\}$ has the symmetry property
\be \label{ecurlysym}
\{\Phi_1 \Phi_2\cdots \Phi_{i-1}\Phi_{i+1} \Phi_i\Phi_{i+2} \cdots \Phi_N\}
=(-1)^{\gamma_i \gamma_{i+1}} \{\Phi_1\Phi_2\cdots \Phi_N\}\, ,
\ee
where $\gamma_i$ is the grassmannality of $|\Phi_i\rangle$.
Using this symmetry property we can bring all the NS sector states at the beginning of the
set of $|\Phi_i\rangle$'s. Thus it will be enough to define $\{\Phi_1\cdots \Phi_N\}$
for such an arrangement of the $|\Phi_i\rangle$'s.
\item For $|\Phi_1\rangle,\cdots |\Phi_{\m}\rangle\in\HH_{NS}$ and 
$|\Phi_{\m+1}\rangle,\cdots |\Phi_{\m+\n}\rangle\in\HH_{R}$, we define
\be \label{edefcurly}
\{\Phi_1\cdots \Phi_{\m+\n}\}= \sum_{g=0}^\infty (g_s)^{2g} 
\int_{\RR_{g,\m,\n}} \Omega^{(g,\m,\n)}_{6g-6+2\m+2\n}(|\Phi_1\rangle,\cdots |\Phi_{\m+\n}
\rangle)\, .
\ee
Note that the property \refb{ecurlysym} under the exchange of an NS sector state with
an R sector state is part of the definition of $ \{\Phi_1\Phi_2\cdots \Phi_N\}$, whereas 
the same property under the exchange of two NS sector states or two R sector states
follows from the property \refb{exchange} of $\Omega^{(g,\m,\n)}_p$ and the fact that 
$\RR_{g,m,n}$
is symmetric under the exchange of the NS-punctures and also under the exchange of
the R-punctures.
\end{enumerate}
It follows from the property of $\Omega^{(g,m,n)}_p$
that if the set $|\Phi_1\rangle,\cdots |\Phi_N\rangle$ contains $m$ NS sector and $n$ R-sector
states then in order to get non-vanishing result for $\{\Phi_1\cdots \Phi_N\}$ 
we must have $\sum_{i=1}^N n_i = 2N$ and $\sum_{i=1}^N q_i = -m - n/2$,
where $(n_i, q_i)$ are the ghost  and picture numbers of $|\Phi_i\rangle$. 
As a consequence of \refb{eomega} and \refb{eboundary},
\refb{efactor}, \refb{efactoraa}
we have the following important
identity
\ben \label{eimpid}
&&  \sum_{i=1}^N (-1)^{\gamma_1+\cdots \gamma_{i-1}}\{\Phi_1\cdots \Phi_{i-1} (Q_B \Phi_i)
\Phi_{i+1} \cdots \Phi_N\} \nonumber \\
&=& -  
{1\over 2} \sum_{\ell,k\ge 0\atop \ell+k=N} \sum_{\{i_a;a=1,\cdots \ell\}, \{j_b;b=1,\cdots k\}\atop
\{i_a\}\cup \{j_b\} = \{1,\cdots N\}
}\sigma(\{i_a\}, \{j_b\})
\{\Phi_{i_1} \cdots \Phi_{i_\ell} \vp_r\} \{ (\GG \vp^r) \Phi_{j_1} \cdots \Phi_{j_k}\} 
\, %\nonumber \\
\een
where $\sigma(\{i_a\}, \{j_b\})$ is the sign that one picks up while rearranging
$b_0^-,\Phi_1,\cdots \Phi_N$ to\break \noindent
$\Phi_{i_1},\cdots \Phi_{i_\ell}, b_0^-, \Phi_{j_1},\cdots \Phi_{j_k}$
and $\{|\vp_r\rangle\}$ and
$\{|\vp^r\rangle\}$ are a set of dual basis of $\HH_{T}$
satisfying
\be \label{ebasisbb}
\langle \vp^r | c_0^-|\vp_s\rangle = \delta^r{}_s \qquad \Leftrightarrow \qquad
\langle \vp_s | c_0^-|\vp^r\rangle = \delta^r{}_s  \, ,
\ee
and the completeness relation
\be \label{ecomplete}
|\vp_r\rangle \langle \vp^r| = |\vp^r\rangle \langle \vp_r|=b_0^-\, .
\ee
Note the use of the symbol $\GG$ defined in \refb{edefgg} -- it is identity if 
$|\vp_r\rangle\in\HH_{NS}$ and $\XX_0$ if $|\vp_r\rangle\in\HH_{R}$.
If $\m$ of the $|\Phi_i\rangle$'s represent NS sector states and $\n=N-\m$
of the $|\Phi_i\rangle$'s represent R sector states then the coefficient of
the $g_s{}^{2g}$ term on
the left hand side of \refb{eimpid} is 
the integral of the left hand side of \refb{eomega} over
appropriate $\RR_{g,m,n}$. On the other hand the coefficient of the $g_s{}^{2g}$ term on
the right hand side of \refb{eimpid}
represents the boundary terms that one obtains by integrating the total derivative term
on the right hand side of \refb{eomega} over $\RR_{g,m,n}$. 
These boundary terms can be evaluated using
\refb{eboundary} and the factorization properties \refb{efactor}, \refb{efactoraa}
yielding the expression given on the right hand side of \refb{eimpid}. Special attention
must be paid to the signs. The overall minus sign on the right hand side of
\refb{eimpid} has its origin in the minus sign on the right hand side of \refb{eboundary}.
The  $\sigma(\{i_r\}, \{j_s\})$ factor in \refb{eimpid} represents the product
$\tilde\sigma_1\tilde \sigma_5$ or $\hat\sigma_1\hat\sigma_5$ in \refb{efactor},
\refb{efactoraa}. The $\tilde\sigma_3,\tilde\sigma_4$ factors
in \refb{efactor} and $\hat\sigma_3,\hat\sigma_4$ factors in
\refb{efactoraa} are unity since the degrees $p_1$ and $p_2$ of the differential forms are
even.
Finally the
$\tilde\sigma_2$ factor in \refb{efactor} and $\hat\sigma_2$ factor in \refb{efactoraa}
are not required in \refb{eimpid} since the $|\tilde\vp_r\rangle$ and 
$\XX_0|\hat\vp^r\rangle$ factors
which were in the `incorrect positions' in these equations requiring this sign have been
moved back to the `correct position'  sitting next to each other in \refb{eimpid}.

Next we introduce a multilinear function $|[\Phi_2\cdots \Phi_N]\rangle\in\HH_T$ 
of $(N-1)$ variables
$|\Phi_2\rangle,\cdots |\Phi_N\rangle\in \HH_T$, defined via the relations
\be \label{edefsquare}
\langle \Phi_1| c_0^- |[\Phi_2\cdots \Phi_N]\rangle = \{\Phi_1\cdots \Phi_N\}
\ee
for all $|\Phi_1\rangle\in\HH_T$. 
Here $\langle A|B\rangle$ denotes the BPZ inner product.
As in \cite{1411.7478} we have dropped the ket symbol
$|~\rangle$ from the states when they appear in the argument of
$\{~\}$ or $[~]$. We shall also drop the ket symbol from $|[\Phi_2\cdots \Phi_N]\rangle$
except in inner products. If the set $|\Phi_1\rangle,\cdots |\Phi_N\rangle$   
contains $m$ NS sector and $n$ R-sector
states then 
$[\Phi_2\cdots \Phi_N]$ has ghost number equal to $3 +\sum_{i=2}^N n_i-2(N-1)$ and
picture number equal to $m + [n/2] + \sum_{i=2}^N q_i-1$  where $[n/2]$ denotes the largest
integer $\le n/2$. In particular if all the NS sector states are in the $-1$ picture
and all the R-sector states are in the $-1/2$ picture then the picture number of
$[\Phi_2\cdots \Phi_N]$ is $-1$ if $n$ is even and $-3/2$ if $n$ is odd.

Eq.\refb{ecurlysym} can now be translated to the identity
\be \label{esymmetry}
[\Phi_2\cdots \Phi_{i-1}\Phi_{i+1} \Phi_i\Phi_{i+2} \cdots \Phi_N]
=(-1)^{\gamma_i \gamma_{i+1}} [\Phi_2\cdots \Phi_N] \, .
\ee
Furthermore \refb{eimpid} tells us that for $N\ge 1$,\footnote{Note that 
inside $[\cdots]$ in the first term of \refb{emain}
the first argument is $\Phi_2$ and hence there are
only $N-1$ arguments. Thus for $N=1$ we have the equation $Q_B[] +[\GG[]]=0$.
}
\ben \label{emain}
&& Q_B[\Phi_2\cdots \Phi_N] + \sum_{i=2}^N (-1)^{\gamma_2+\cdots 
\gamma_{i-1}}[\Phi_2\cdots \Phi_{i-1} (Q_B \Phi_i)
\Phi_{i+1} \cdots \Phi_N] \nonumber \\
&=& -  \sum_{\ell,k\ge 0\atop \ell+k=N-1} \sum_{\{i_a;a=1,\cdots \ell\}, \{j_b;b=1,\cdots k\}\atop
\{i_a\}\cup \{j_b\} = \{2,\cdots N\}
}\sigma(\{i_a\}, \{j_b\})\, 
[\Phi_{i_1} \cdots \Phi_{i_\ell} \GG\, [\Phi_{j_1} \cdots \Phi_{j_k}]] 
\een
where in
the last term the sum runs over all possible ways of splitting the
set $\{2,\cdots N\}$ into the set $\{i_a\}$ and the set $\{j_b\}$. 
$\sigma(\{i_a\}, \{j_b\})$ is the sign that one picks up while rearranging
$b_0^-,\Phi_2,\cdots \Phi_N$ to 
$\Phi_{i_1},\cdots \Phi_{i_\ell}, b_0^-, \Phi_{j_1},\cdots \Phi_{j_k}$.
The inner product of \refb{emain} with an arbitrary state $\langle\Phi_1|c_0^-$ is 
given by $(-1)^{\gamma_1}$ times  \refb{eimpid}. For the left hand sides the equality is
obvious. For the right hand side, we note that in \refb{eimpid} we have two
kinds of contributions: $\Phi_1$ can either be inside the first curly bracket or be
inside the second curly bracket. These two contributions
are identical due to the identity
\be \label{eneweq2a}
\{\Phi_1\cdots \Phi_k \vp_r\} \{(\GG\vp^r)\wt\Phi_1\cdots \wt\Phi_\ell\}
= (-1)^{\gamma+\tilde\gamma+\gamma\tilde\gamma}
\{\wt\Phi_1\cdots \wt\Phi_\ell \vp_r\} \{(\GG\vp^r)\Phi_1\cdots \Phi_k\}\, ,
\ee
which we shall prove shortly.
Here $\gamma$ is the total grassmannality of $|\Phi_1\rangle,\cdots|\Phi_k\rangle$
and $\tilde\gamma$ is the total grassmannality of 
$|\tilde\Phi_1\rangle,\cdots|\tilde\Phi_\ell\rangle$.
Assuming this to be the case, we can only keep the terms on the right hand side of
\refb{eimpid} where
$\Phi_1$ is inside the first curly bracket and multiply the result by 2. After being multiplied
by $(-1)^{\gamma_1}$, this reproduces
the inner product of $\langle\Phi_1|c_0^-$ with the right hand side of \refb{emain}.

Let us now prove \refb{eneweq2a}. 
First we switch the order of the two terms on the right hand side of \refb{eneweq2a}
to express this as
\ben \label{e4.10}
&&\{\Phi_1\cdots \Phi_k \vp_r\} \{(\GG\vp^r)\wt\Phi_1\cdots \wt\Phi_\ell\}
\nonumber \\ &=& 
(-1)^{\gamma+\tilde\gamma+\gamma\tilde\gamma+ (\gamma_{\vp_r}+\tilde\gamma)
(\gamma_{\vp_r}+1+\gamma)}
 \{(\GG\vp^r)\Phi_1\cdots \Phi_k\} \{\wt\Phi_1\cdots \wt\Phi_\ell \vp_r\}\, ,
\een
where $\gamma_{\vp_r}$, $\gamma_{\vp^r}$ are the grassmannalities of
$|\vp_r\rangle$ and $|\vp^r\rangle$ and 
we have used $\gamma_{\vp^r}=\gamma_{\vp_r}+1$ mod 2. The latter relation follows
from \refb{ebasisbb}.
Using \refb{edefsquare},
\refb{ecurlysym} we can express \refb{e4.10} as
\ben \label{exy0}
&& \langle \vp_r|c_0^-| [\Phi_1\cdots \Phi_k]\rangle \langle \GG\vp^r|c_0^-|
[\wt\Phi_1\cdots \wt\Phi_\ell]\rangle\nonumber \\
&=& 
 \langle \GG\vp^r|c_0^-|
[\Phi_1\cdots \Phi_k]\rangle \langle \vp_r|c_0^-| [\wt\Phi_1\cdots \wt\Phi_\ell]\rangle
(-1)^{\gamma+\tilde\gamma+\gamma\tilde\gamma
+(\gamma_{\vp_r}+\tilde\gamma)
(\gamma_{\vp_r}+1+\gamma) +\gamma_{\vp_r}(\gamma+\tilde\gamma)}\, .
\een
Now we have 
\be \label{exy1}
\langle A|c_0^-|B\rangle=(-1)^{\gamma_A+\gamma_B+\gamma_A\gamma_B+1}
\langle B|c_0^-|A\rangle, \quad \langle\GG A|c_0^-|B\rangle
=(-1)^{\gamma_A+\gamma_B+\gamma_A\gamma_B+1} \langle B|c_0^- \GG|A\rangle\, .
\ee
Applying the first equation on the first term 
on the left hand side of \refb{exy0} and the second equation on the first term
on the right hand side of \refb{exy0}, and noting that the grassmannality of
$[\Phi_1\cdots \Phi_k]$ is $\gamma+1$ mod 2,  we can express \refb{exy0} as
\ben \label{exy01}
&& \langle  [\Phi_1\cdots \Phi_k]|c_0^-|\vp_r\rangle \langle \GG\vp^r|c_0^-|
[\wt\Phi_1\cdots \wt\Phi_\ell]\rangle \nonumber \\
&=& (-1)^{\gamma+\tilde\gamma+\gamma\tilde\gamma
+(\gamma_{\vp_r}+\tilde\gamma)
(\gamma_{\vp_r}+1+\gamma) +\gamma_{\vp_r}(\gamma+\tilde\gamma)
+ \gamma_{\vp_r}\gamma +(\gamma_{\vp_r}+1)
\gamma}  \nonumber \\ && \times
 \langle[\Phi_1\cdots \Phi_k] |c_0^-
\GG|\vp^r \rangle 
\langle \vp_r|c_0^-| [\wt\Phi_1\cdots \wt\Phi_\ell]\rangle\, .
\een
Using the completeness relations \refb{ecomplete}, 
the fact that $[\GG,c_0^-]$ vanishes when sandwiched between states annihilated
by $b_0^-$, and simplifying the exponent of $(-1)$,
\refb{exy01} reduces to
\be
\langle [\Phi_1\cdots \Phi_k]| c_0^- \GG | [\wt\Phi_1\cdots \wt\Phi_\ell]\rangle
= \langle [\Phi_1\cdots \Phi_k]| c_0^- \GG | [\wt\Phi_1\cdots \wt\Phi_\ell]\rangle\, ,
\ee
which is an identity. This in turn proves
\refb{eneweq2a}.

\subsection{The equation of motion and its gauge invariance}

A general string field configuration is taken to be an element $|\Psi\rangle$ 
of $\HH_{(-1)}\oplus
\HH_{(-1/2)}$ of ghost number 2, with the component along $\HH_{(-1)}$ representing 
the bosonic
fields and the component along $\HH_{(-1/2)}$ representing the fermionic fields.
Thus $|\Psi\rangle$ is grassmann even.
The equations of motion for $|\Psi\rangle$ in the 1PI effective heterotic 
string field theory 
is taken to be\footnote{Alternatively we could take the string field to be an element
$|\wt\Psi\rangle\in\HH_{(-1)}\oplus
\HH_{(-3/2)}$ of ghost number 2, and write the equation of motion as
$Q_B|\wt\Psi\rangle + \sum_{n=1}^\infty {1\over (n-1)!} [(\GG\wt\Psi)^{n-1}]=0$.
$|\wt\Psi\rangle$ and $|\Psi\rangle$ will be related as $|\Psi\rangle=\GG|\wt\Psi\rangle$.
Since the cohomology of $Q_B$ in picture numbers $-1/2$ and $-3/2$ 
coincide\cite{9711087},
this will give a sensible set of equations of motion.
We shall not explore this in detail, but the reader will find some 
related comments at the end of \S\ref{smatrix}. \label{f5}}
\be \label{eeom}
|\EE\rangle=0, \qquad |\EE\rangle\equiv 
Q_B|\Psi\rangle + \sum_{n=1}^\infty {1\over (n-1)!} \GG[\Psi^{n-1}]\, .
\ee
Note that $Q_B|\Psi\rangle$ is an element of $\HH_{(-1)}+\HH_{(-1/2)}$ of ghost number
3 whereas $[\Psi^{n-1}]$ is an element of $\HH_{(-1)}+\HH_{(-3/2)}$ of ghost number
3. The operation of $\GG$ is essential to map the latter to an element of 
$\HH_{(-1)}+\HH_{(-1/2)}$. 
The infinitesimal gauge transformation is generated by an element $|\Lambda\rangle$ 
of $\HH_{(-1)}\oplus
\HH_{(-1/2)}$ of ghost number 1. $|\Lambda\rangle$ is grassmann odd.
The gauge transformation law of $|\Psi\rangle$ is
\be  \label{egauge}
|\delta\Psi\rangle = Q_B|\Lambda\rangle + \sum_{n=0}^\infty {1\over n!} 
\GG[\Psi^n \Lambda]
\ee
Again the operation of $\GG$ is crucial in bringing $[\Psi^n \Lambda]$ which is an element
of $\HH_{(-1)}+\HH_{(-3/2)}$ to an element of $\HH_{(-1)}\oplus
\HH_{(-1/2)}$.

We shall now show that the equations of motion are gauge covariant, i.e.\ if $|\Psi\rangle$
satisfies equations of motion then its gauge transform also satisfies equations of motion.
Taking the gauge variation of \refb{eeom} gives
\be \label{exy5}
|\delta \EE\rangle =
Q_B|\delta\Psi\rangle + \sum_{n=1}^\infty {1\over (n-1)!} \GG[\Psi^{n-1} \delta\Psi]\, .
\ee
Our goal is to show that $|\delta \EE\rangle$ vanishes when $|\EE\rangle$ vanishes. Now
using \refb{egauge} we can express \refb{exy5} as
\be
|\delta \EE\rangle = \sum_{n=0}^\infty {1\over n!} 
Q_B \GG[\Psi^n \Lambda] + 
\sum_{n=1}^\infty {1\over (n-1)!} \GG[\Psi^{n-1} Q_B\Lambda]
+ \sum_{n=1}^\infty {1\over (n-1)!}  \sum_{m=0}^\infty {1\over m!} 
\GG\left[\Psi^{n-1} 
\GG[\Psi^m \Lambda]\right]\, .
\ee
We now manipulate the first term on the right hand side using $[Q_B,\GG]=0$
and \refb{emain}. Since $|\Psi\rangle$ is grassmann even and $|\Lambda\rangle$
is grassmann odd, we get
\ben
|\delta \EE\rangle &=& -\sum_{n=1}^\infty {1\over (n-1)!} 
\GG[\Psi^{n-1} (Q_B\Psi)  \Lambda] - \sum_{n=0}^\infty {1\over n!} 
\GG[\Psi^{n} Q_B\Lambda] - \sum_{p=0}^\infty {1\over p!} \sum_{m=0}^\infty {1\over m!}
\GG[\Psi^p \GG[\Psi^m\Lambda]] \nonumber \\ && + 
\sum_{p=0}^\infty {1\over p!} \sum_{m=0}^\infty {1\over m!}
\GG[\Psi^p \Lambda\GG[\Psi^m]]
 \nonumber \\
&& + 
\sum_{n=1}^\infty {1\over (n-1)!} \GG[\Psi^{n-1} Q_B\Lambda]
+ \sum_{n=1}^\infty {1\over (n-1)!}  \sum_{m=0}^\infty {1\over m!} 
\GG\left[\Psi^{n-1} 
\GG[\Psi^m \Lambda]\right]\, .
\een
The first and the fourth term on the right hand side cancel using the equations
of motion \refb{eeom}. The second and fifth terms cancel and the third and the
sixth terms cancel. Thus we get
\be
|\delta \EE\rangle = 0\, .
\ee
This proves that the equations of motion transform covariantly under gauge 
transformations.

Note that if we restrict to the states in the NS sector then $\GG$ can be replaced by
the identity operator and the equations of motion reduce to those which were derived from
the 1PI action in \cite{1411.7478}. 
However once we include the R-sector states there is no
fully satisfactory action from which the equations of motion \refb{eeom}
can be derived. More discussion on this can be found in \S\ref{smatrix}.

\subsection{Auxiliary action and S-matrix elements} \label{smatrix}

We shall now argue that the tree level Green's functions computed from the 1PI
effective theory described above reproduces the 
off-shell amplitudes described in \S\ref{soff}. In that case the S-matrix elements
computed via LSZ prescription from these two approaches would also agree. 
Although it is in principle possible to compute the tree level S-matrix from the equations 
of motion directly we shall take a short-cut by 
using an action {\it with additional states} from which the equations of motion can be derived.
For this we introduce a new set of fields $|\wt\Psi\rangle\in \HH_{(-1)}\oplus\HH_{(-3/2)}$
of ghost number 2 
and consider the
action
\be \label{eact}
S = g_s{}^{-2}\left[
-{1\over 2} \langle\wt\Psi |c_0^- Q_B \GG |\wt\Psi\rangle 
+ \langle\wt\Psi |c_0^- Q_B |\Psi\rangle + \sum_{n=0}^\infty {1\over n!} \{ \Psi^n\}
\right]\, .
\ee
The equation of motion for $|\wt\Psi\rangle$ derived from \refb{eact} is
\be \label{e01}
Q_B (|\Psi\rangle - \GG|\wt\Psi\rangle) = 0\, .
\ee
On the other hand
the equation of motion of $|\Psi\rangle$ is
\be \label{e02}
Q_B |\wt\Psi\rangle + \sum_{n=1}^\infty {1\over (n-1)!} [\Psi^{n-1}] = 0\, .
\ee
Applying $\GG$ on \refb{e02} and using \refb{e01} we recover the equation of
motion \refb{eeom} of $|\Psi\rangle$. It is easy to see that the action 
\refb{eact} is invariant under the infinitesimal gauge transformation \refb{egauge} if we also
transform $|\wt\Psi\rangle$ as
\be
|\delta\wt\Psi\rangle = Q_B|\wt\Lambda\rangle + \sum_{n=0}^\infty {1\over n!} 
[\Psi^n \Lambda]\, ,
\ee
where $|\wt\Lambda\rangle\in \HH_{(-1)}\oplus\HH_{(-3/2)}$ and carries ghost number 1.

We can now gauge fix the theory in the Siegel gauge $b_0^+|\Psi\rangle=0$,
$b_0^+|\wt\Psi\rangle=0$ in which case the kinetic term in the $(|\wt\Psi\rangle, |\Psi\rangle)$
space is proportional to
\be
c_0^- c_0^+ L_0^+ \pmatrix{-\GG & 1\cr 1 & 0}\, ,
\ee
leading to the propagator
\be 
b_0^+ b_0^- (L_0^+)^{-1} \delta_{L_0,\bar L_0}\pmatrix{0 & 1\cr 1 & \GG}\, .
\ee
Since the interaction terms involve only the field $|\Psi\rangle$, only the $|\Psi\rangle$
propagator is relevant for computing the Green's functions in the Siegel gauge with external legs
truncated.
This is proportional to $b_0^- b_0^+ (L_0^+)^{-1}\delta_{L_0,\bar L_0}\GG$
which is precisely the propagator used in
the analysis of \S\ref{soff}.
Standard argument then shows that the
off-shell Green's functions with external tree
level propagator truncated, and only
$|\Psi\rangle$ as external states,
coincide with the off-shell amplitudes described in \S\ref{soff}, with
the 1PI contribution reproducing the part of the integration cycle that is described by
$\RR_{g,m,n}$ and the one particle reducible (1PR) 
contributions reproducing the rest of the components of the integration
cycle. Together they describe the full integration cycle whose projection on
the base covers the whole of $\MM_{g,m,n}$. Thus the S-matrix elements computed
from these Green's functions will also agree with the ones computed from the
off-shell amplitudes described in \S\ref{soff}. 

Could we use the action \refb{eact} for  defining the 1PI effective theory? The problem 
with this is that the equations of motion \refb{e01}, \refb{e02} have more
classical solutions than the ones expected in string theory. For example at the linearized
level we can 
consider solutions with $|\wt\Psi\rangle=0$, 
$Q_B|\Psi\rangle=0$, and independently another set of solutions for which 
$|\Psi\rangle=0$, $Q_B|\wt\Psi\rangle=0$. This will double the number of physical
states. 
This
does not make any difference as long as we are using this action to 
compute tree level S-matrix elements
with external $|\Psi\rangle$ states (which could in principle be computed just from
the equations of motion \refb{eeom} of $|\Psi\rangle$), 
but due to these extra states the action
\refb{eact} cannot be regarded as the fundamental action for describing the 1PI
effective string field theory. We could try to remove the extra states by adding a 
constraint $|\Psi\rangle = \GG|\wt\Psi\rangle$. Now \refb{e01} holds automatically and
\refb{e02} reproduces the equations of motion described in
footnote \ref{f5}, but this constraint has to be imposed externally
and does not follow from the action. On the other hand if we use this constraint
to eliminate $|\Psi\rangle$ from the action \refb{eact} and treat $|\wt\Psi\rangle$ as
independent field variables, then 
we arrive at the action
\be \label{eaction32}
g_s{}^{-2} \left[ {1\over 2} \langle \wt\Psi | c_0^- Q_B \GG |\wt\Psi\rangle +
\sum_{n=1}^\infty {1\over n!} \{ (\GG\wt\Psi)^n\}\right]\, .
\ee
Now
the kinetic operator becomes proportional to
$c_0^- Q_B \GG$ which may have additional zeroes from the kernel of $\GG$ and hence
the spectrum of physical states in this theory will again differ from the expected
spectrum of string theory.

\subsection{Generalizaton to type II string theories}

Generalization of the above analysis to type II string theories is straightforward.
The only difference is that the string field is now taken to be a
grassmann even  element
of $\HH_{(-1,-1)}\oplus \HH_{(-1,-1/2)}\oplus \HH_{(-1/2,-1)}\oplus 
\HH_{(-1/2,-1/2)}$ carrying ghost number 2. 
However the use of the symbol $\GG$ -- now defined as in
\refb{edefggii} -- ensures that all
the formul\ae\ derived for the heterotic string theory continue to be valid for type
II string theories. The analysis of \S\ref{smatrix} can also be extended to this case by
taking
$|\wt\Psi\rangle\in \HH_{(-1,-1)}\oplus\HH_{(-1,-3/2)}
\oplus\HH_{(-3/2,-1)}\oplus\HH_{(-3/2,-3/2)}$. 

\sectiono{Effect of changing the local coordinates and/or PCO locations} \label{seffect}

We now study the effect of changing the choice of
local coordinates and/or the locations of the PCO's on the 1PI effective theory
following \cite{9301097}. 
A change of
this form will correspond to a new choice of the $(6g-6+2m+2n)$ dimensional 
subspaces $\RR_{g,m,n}$ in $\wt\PP_{g,m,n}$
satisfying \refb{eboundary}. Let us denote them by $\RR'_{g,m,n}$. We shall consider
infinitesimal deformations so that $\RR_{g,m,n}$ and $\RR'_{g,m,n}$ are close in
$\wt\PP_{g,m,n}$ and denote the corresponding change in $|\EE\rangle$ defined in \refb{eeom}
by $|\wh\delta \EE\rangle$. Our goal will be to show that there is a possible field redefinition
$|\Psi\rangle \to |\Psi\rangle +|\tilde\delta\Psi\rangle$ such that the change $|\tilde\delta\EE\rangle$
in $|\EE\rangle$ induced by this field redefinition reproduces $|\wh\delta\EE\rangle$ upon using equations
of motion, i.e.\
\be \label{edetde}
|\wh\delta \EE\rangle -|\tilde\delta\EE\rangle = 0
\ee
when $|\EE\rangle=0$. This will imply that the effect of the change in local coordinates
/ PCO  locations can be compensated by a field redefinition.

To proceed, let us decompose the string field $|\Psi\rangle$ into its NS and R part:
\be
|\Psi\rangle = |\Psi_{NS}\rangle +|\Psi_R\rangle
\ee
so that we can write
\be
|\EE\rangle = Q_B(|\Psi_{NS}\rangle +|\Psi_R\rangle)
+ \sum_{m,n=0}^\infty {1\over 
m! n!} \GG[(\Psi_{NS})^m (\Psi_R)^n]\, .
\ee
This gives, using \refb{edefsquare},
\be \label{e5.4}
\langle \Phi|c_0^-|\EE\rangle = \langle \Phi|c_0^-Q_B(|\Psi_{NS}\rangle +|\Psi_R\rangle)
+ \sum_{m,n=0}^\infty {1\over 
m! n!} \{(\GG\Phi)(\Psi_{NS})^m (\Psi_R)^n\}\, ,
\ee
for any state $|\Phi\rangle\in \HH_{(-1)}\oplus\HH_{(-3/2)}$ of ghost number 2. 
We shall take $|\Phi\rangle$ to be grassmann even for convenience, but the analysis can be repeated for
grassmann odd $|\Phi\rangle$ as well. Decomposing
$|\Phi\rangle$ as $|\Phi_{NS}\rangle+|\Phi_R\rangle$ we get, using \refb{edefcurly} and \refb{e5.4}
\ben \label{edeltaactionpre}
&& \langle \Phi|c_0^-|\wh\delta \EE\rangle \nonumber \\
&=& \sum_{g=0}^\infty  g_s{}^{2g}  \sum_{m,n=0}^\infty {1\over 
m! n!} \,
\left[\left(\int_{\RR'_{g,m+1,n}} - \int_{\RR_{g,m+1,n}}\right)
\, \Omega^{(g,m+1,n)}_{6g-6+2m+2n+2}(\GG|\Phi_{NS}\rangle,|\Psi_{NS}\rangle^{\otimes m},
|\Psi_R\rangle^{\otimes n})
\right] \nonumber \\
&+& \sum_{g=0}^\infty  g_s{}^{2g}  \sum_{m,n=0}^\infty {1\over 
m! n!} \,
\left[\left(\int_{\RR'_{g,m,n+1}} - \int_{\RR_{g,m,n+1}}\right)
\, \Omega^{(g,m,n+1)}_{6g-6+2m+2n+2}(|\Psi_{NS}\rangle^{\otimes m},\GG|\Phi_{R}\rangle,
|\Psi_R\rangle^{\otimes n})
\right] \nonumber \\
\een
where $|\Psi\rangle^{\otimes n}$ denotes that there are $n$ entries of
$|\Psi\rangle$ in the argument.

Let us first focus on the part involving $|\Phi_{NS}\rangle$ so that we only have the first 
term on the right hand side of \refb{edeltaactionpre}. 
Let $\wh U_{g,m,n}$ be an infinitesimal vector field that takes a point in $\RR_{g,m,n}$ to a
neighbouring point in $\RR'_{g,m,n}$.  $\wh U_{g,m,n}$ is defined up to addition
of tangent vectors of $\RR_{g,m,n}$.
In this case the part of \refb{edeltaactionpre} involving $|\Phi_{NS}\rangle$ 
can be expressed as\cite{9301097,1411.7478}
\ben \label{edeltaaction}
&& \langle \Phi_{NS}|c_0^-|\wh\delta \EE\rangle \nonumber \\
&=& \sum_{g=0}^\infty  g_s{}^{2g}  
\sum_{m,n=0}^\infty {1\over m! n!} \, 
\left[\int_{\RR_{g,m+1,n}}
\, d\Omega^{(g,m+1,n)}_{6g-6+2m+2n+2}[\wh U_{g,m+1,n}] 
(\GG|\Phi_{NS}\rangle, |\Psi_{NS}\rangle^{\otimes m}, 
|\Psi_R\rangle^{\otimes n})  \right.  \nonumber \\ &&
 \left. +
\int_{\p \RR_{g,m+1,n}} \Omega^{(g,m+1,n)}_{6g-6+2m+2n+2}[\wh U_{g,m+1,n}] 
(\GG|\Phi_{NS}\rangle, |\Psi_{NS}\rangle^{\otimes m}, |\Psi_R\rangle^{\otimes n}) 
\right] \, ,
\een
where for any $p$-form $\omega_p$, 
$\omega_p[\wh U]$ denotes the contraction of 
$\omega_p$ with the vector field $\wh U$:
\be
\omega_{i_1\cdots i_p}dy^{i_1} \wedge \cdots \wedge dy^{i_p}[\wh U] \equiv \wh U^{i_1} 
\omega_{i_1 i_2\cdots i_p}dy^{i_2} \wedge \cdots \wedge dy^{i_p}\, .
\ee
An intuitive understanding of \refb{edeltaaction} can be found in Fig.~1 of \cite{1411.7478}.
We  manipulate the first term on the right hand side of \refb{edeltaaction}
using \refb{eomega} and the second term using
\refb{efactor} and \refb{efactoraa}. The second term can be further simplified by noting that
on $\p\RR_{g,m+1,n}$ -- which can be regarded as the result
 of plumbing fixture of two 1PI Riemann surfaces with $s=0$ --
the state $\GG|\Phi_{NS}\rangle$ can be inserted either on the first surface
or on the second surface. Since these contributions are equal, we shall insert
$\GG|\Phi_{NS}\rangle$ on the first surface and multiply the result by a factor
of two. Furthermore on $\p\RR_{g,m+1,n}$
the vector field $\wh U_{g,m,n}$ reduces to the sum of two
vector fields labelling the deformations of the choice of local coordinates and
PCO locations on the two components that are glued to produce $\p\RR_{g,m+1,n}$.
This gives
\ben \label{edeltaactionbb}
&& \langle \Phi_{NS}|c_0^-|\wh\delta \EE\rangle \nonumber \\
&=& \sum_{g=0}^\infty  g_s{}^{2g}   \left[
-\sum_{m,n=0}^\infty {1\over m! n!} \, 
\int_{\RR_{g,m+1,n}}
\, \Omega^{(g,m+1,n)}_{6g-6+2m+2n+3}[\wh U_{g,m+1,n}] 
(Q_B\GG|\Phi_{NS}\rangle, |\Psi_{NS}\rangle^{\otimes m}, 
|\Psi_R\rangle^{\otimes n})  \right. \nonumber \\ &-&
\sum_{m=1}^\infty \sum_{n=0}^\infty {1\over (m-1)! n!} \, 
\int_{\RR_{g,m+1,n}}
\, \Omega^{(g,m+1,n)}_{6g-6+2m+2n+3}[\wh U_{g,m+1,n}] 
(\GG|\Phi_{NS}\rangle, Q_B|\Psi_{NS}\rangle, |\Psi_{NS}\rangle^{\otimes {(m-1)}}, 
|\Psi_R\rangle^{\otimes n})  \nonumber \\ &-&
\left. 
\sum_{m=0}^\infty \sum_{n=1}^\infty {1\over m! (n-1)!} \, 
\int_{\RR_{g,m+1,n}}
\, \Omega^{(g,m+1,n)}_{6g-6+2m+2n+3}[\wh U_{g,m+1,n}] 
(\GG|\Phi_{NS}\rangle, |\Psi_{NS}\rangle^{\otimes m}, 
Q_B|\Psi_R\rangle, |\Psi_R\rangle^{\otimes {(n-1)}}) \right] \nonumber \\ &-&
%\left. +
%\sum_{m,n=0}^\infty {1\over m! n!} \, 
%\int_{\p \RR_{g,m+1,n}} \Omega^{(g,m+1,n)}_{6g-6+2m+2n+2}[\wh U_{g,m+1,n}] 
%(\GG|\Phi_{NS}\rangle, |\Psi_{NS}\rangle^{\otimes m}, |\Psi_R\rangle^{\otimes n}) 
%\right]
%\nonumber \\ && 
\sum_{g_1,g_2} g_s{}^{2(g_1+g_2)} \, 
\sum_{m_1,m_2,n_1,n_2\ge 0}
{1\over m_1! m_2! n_1! n_2!}
\int_{\{\RR_{g_1,m_1+2,n_1}, \RR_{g_2,m_2+1,n_2}\}} \nonumber \\ &&
\Omega^{(g_1+g_2,m_1+m_2+1,n_1+n_2)}_{6(g_1+g_2)
-6+2(m_1+m_2+n_1+n_2+1)}
[\wh U_{g_1,m_1+2,n_1} + \wh U_{g_2,m_2+1,n_2}] 
(\GG|\Phi_{NS}\rangle, |\Psi_{NS}\rangle^{\otimes (m_1+m_2)}, |\Psi_R\rangle^{\otimes (n_1+n_2)}) 
\nonumber \\
&-& \sum_{g_1,g_2} g_s{}^{2(g_1+g_2)} 
\sum_{m_1,m_2,n_1,n_2\ge 0}
{1\over m_1! m_2! n_1! n_2!}
\int_{\{\RR_{g_1,m_1+1,n_1+1};\RR_{g_2,m_2,n_2+1}\}} \nonumber \\ &&
\Omega^{(g_1+g_2,m_1+m_2+1,n_1+n_2)}_{6(g_1+g_2)
-6+2(m_1+m_2+n_1+n_2+1)}
[\wh U_{g_1,m_1+1,n_1+1} + \wh U_{g_2,m_2,n_2+1}] 
(\GG|\Phi_{NS}\rangle, |\Psi_{NS}\rangle^{\otimes (m_1+m_2)}, |\Psi_R\rangle^{\otimes (n_1+n_2)}) 
\nonumber \\
&\equiv& I_1+I_2+I_3+I_4+I_5\, ,
\een
where $I_1,\cdots I_5$ denote the five terms appearing on the right hand side.
In the expressions for $I_4$ and $I_5$ it is understood that $\GG|\Phi_{NS}\rangle$,
$m_1$ of the $|\Psi_{NS}\rangle$'s and $n_1$ of the $|\Psi_R\rangle$'s are inserted
on the first Riemann surface and 
$m_2$ of the $|\Psi_{NS}\rangle$'s and $n_2$ of the $|\Psi_R\rangle$'s are inserted
on the second Riemann surface. In the first three terms 
the minus signs have their origin in the $(-1)^p$ factor 
in \refb{eomega}. In the last two terms the
minus signs come from the application of \refb{eboundary}.
The terms involving $|\Phi_R\rangle$
can be analyzed in an identical manner, with the minus signs remaining the same.

We shall now show  that the change in $|\EE\rangle$ given in \refb{edeltaactionbb}
together with its counterpart involving $|\Phi_R\rangle$
can be regarded as the result of a redefinition of the field $|\Psi\rangle$ to
$|\Psi\rangle+|\tilde\delta \Psi\rangle$ in the sense of
\refb{edetde} if we  take $|\tilde\delta\Psi\rangle$ to be of the form
\ben \label{edeltafield}
&& \langle \phi| c_0^- |\tilde\delta \Psi\rangle \nonumber \\
&=& - \sum_{g=0}^\infty g_s{}^{2g} \, \sum_{m,n=0}^\infty  {1\over m!n!} \, 
\int_{\RR_{g,m+1,n}} \, \Omega^{(g,m+1,n)}_{6g-5+2m+2n+2}[\wh U_{g,m+1,n}](\GG|\phi_{NS}
\rangle, 
|\Psi_{NS}\rangle^{\otimes m}, |\Psi_R\rangle^{\otimes n})\nonumber \\ &&
- \sum_{g=0}^\infty g_s{}^{2g} \, \sum_{m,n=0}^\infty  {1\over m!n!} \, 
\int_{\RR_{g,m,n+1}} \, \Omega^{(g,m,n+1)}_{6g-5+2m+2n+2}[\wh U_{g,m,n+1}](
|\Psi_{NS}\rangle^{\otimes m}, \GG|\phi_{R}
\rangle,  |\Psi_R\rangle^{\otimes n})\nonumber \\ \, ,
\een
for any grassmann odd\footnote{The result for grassmann even state can be 
read out by multiplying both sides
of \refb{edeltafield} by a grassmann odd number and moving it through various factors so that it 
multiplies $|\phi\rangle$. 
This gives extra minus signs in both terms on the right hand side of \refb{edeltafield}
since we have to move the grassmann number through the $6g-5+2m+2n+2$ insertions of
$b$-ghost field associated with $  \Omega^{(g,m+1,n)}_{6g-5+2m+2n+2}$.}
state $|\phi\rangle=|\phi_{NS}\rangle + |\phi_R\rangle\in\HH_T$.
Now, using \refb{eeom} we get
\ben
\langle \Phi|c_0^- |\tilde\delta \EE\rangle &=&
\langle \Phi|c_0^- Q_B|\tilde\delta\Psi\rangle + \sum_{n=1}^\infty {1\over (n-1)!} 
\langle \Phi|c_0^- \GG|[\Psi^{n-1} \tilde
\delta\Psi]\rangle\nonumber \\
&=& \langle (Q_B\Phi)| c_0^- |\tilde\delta\Psi\rangle 
+ \sum_{n=1}^\infty {1\over (n-1)!} \{ (\GG\Phi) \Psi^{n-1} \tilde \delta\Psi\}\nonumber \\ 
&=&  \langle (Q_B\Phi)| c_0^- |\tilde\delta\Psi\rangle 
+ \sum_{n=1}^\infty {1\over (n-1)!} \langle [(\GG\Phi) \Psi^{n-1}]| c_0^-
| \tilde \delta\Psi\rangle \, .
\een
Using \refb{edeltafield} we can express this as
\ben
&&- \sum_{g=0}^\infty g_s{}^{2g} \, \sum_{m,n=0}^\infty  {1\over m!n!} \, 
\int_{\RR_{g,m+1,n}} \, \Omega^{(g,m+1,n)}_{6g-5+2m+2n+2}[\wh U_{g,m+1,n}](Q_B
\GG|\Phi_{NS}
\rangle, 
|\Psi_{NS}\rangle^{\otimes m}, |\Psi_R\rangle^{\otimes n})\nonumber \\ &&
- \sum_{g=0}^\infty g_s{}^{2g} \, \sum_{m,n=0}^\infty  {1\over m!n!} \, 
\int_{\RR_{g,m,n+1}} \, \Omega^{(g,m,n+1)}_{6g-5+2m+2n+2}[\wh U_{g,m,n+1}](
|\Psi_{NS}\rangle^{\otimes m}, Q_B\GG|\Phi_{R}
\rangle,  |\Psi_R\rangle^{\otimes n})\nonumber \\ 
&& - \sum_{g=0}^\infty g_s{}^{2g} \, \sum_{p,m,n=0}^\infty  {1\over p!m!n!} \, 
\int_{\RR_{g,m+1,n}} \, \Omega^{(g,m+1,n)}_{6g-5+2m+2n+2}[\wh U_{g,m+1,n}](\GG[(\GG\Phi) \Psi^{p}]_{NS}, 
|\Psi_{NS}\rangle^{\otimes m}, |\Psi_R\rangle^{\otimes n})\nonumber \\ &&
- \sum_{g=0}^\infty g_s{}^{2g} \, \sum_{p,m,n=0}^\infty  {1\over p!m!n!} \, 
\int_{\RR_{g,m,n+1}} \, \Omega^{(g,m,n+1)}_{6g-5+2m+2n+2}[\wh U_{g,m,n+1}](
|\Psi_{NS}\rangle^{\otimes m}, \GG[(\GG\Phi) \Psi^{p}]_{R},  
|\Psi_R\rangle^{\otimes n})\, .\nonumber \\ 
\een
The terms involving $|\Phi_{NS}\rangle$ in the above expression are given by
\ben \label{ejj}
&& \langle \Phi_{NS}|c_0^- |\tilde\delta \EE\rangle \nonumber \\
&=&- \sum_{g=0}^\infty g_s{}^{2g} \, \sum_{m,n=0}^\infty  {1\over m!n!} \, 
\int_{\RR_{g,m+1,n}} \, \Omega^{(g,m+1,n)}_{6g-5+2m+2n+2}[\wh U_{g,m+1,n}](Q_B
\GG|\Phi_{NS}
\rangle, 
|\Psi_{NS}\rangle^{\otimes m}, |\Psi_R\rangle^{\otimes n})\nonumber \\ &&
 - \sum_{g=0}^\infty g_s{}^{2g} \, \sum_{p,m,n=0}^\infty  {1\over p!m!n!} \, 
\int_{\RR_{g,m+1,n}} \, \Omega^{(g,m+1,n)}_{6g-5+2m+2n+2}
[\wh U_{g,m+1,n}](\GG[(\GG\Phi_{NS}) \Psi^{p}]_{NS}, 
|\Psi_{NS}\rangle^{\otimes m}, |\Psi_R\rangle^{\otimes n})\nonumber \\ &&
- \sum_{g=0}^\infty g_s{}^{2g} \, \sum_{p,m,n=0}^\infty  {1\over p!m!n!} \, 
\int_{\RR_{g,m,n+1}} \, \Omega^{(g,m,n+1)}_{6g-5+2m+2n+2}[\wh U_{g,m,n+1}](
|\Psi_{NS}\rangle^{\otimes m}, \GG[(\GG\Phi_{NS}) \Psi^{p}]_{R},  
|\Psi_R\rangle^{\otimes n})\nonumber \\ 
&\equiv & J_1+J_2+J_3\, ,
\een
where $J_1,J_2,J_3$ are the three terms on the right hand side of this expression.
Our goal is to show that \refb{edeltaactionbb} and \refb{ejj} are equal when 
$|\Psi\rangle$ satisfies its equation of motion. First we note that
\be \label{eq1}
I_1 - J_1 = 0\, .
\ee
Next we use \refb{efactor} and \refb{edeltaactionbb}
to write\footnote{As in \cite{1411.7478}, we need slight generalization 
of \refb{efactor} to take into account the contraction of $\Omega^{(g,m,n)}_p$ with
$\wh U_{g,m,n}$.}
\ben \label{enname}
I_4 &=& -
\sum_{g_1,g_2} g_s{}^{2(g_1+g_2)} \, 
\sum_{m_1,m_2,n_1,n_2\ge 0}{1\over m_1! m_2! n_1! n_2!}  \nonumber \\ &&
\times \int_{\RR_{g_1,m_1+2,n_1}}
\Omega^{(g_1,m_1+2,n_1)}_{6g_1
-5+2 (m_1+n_1+2)}
[\wh U_{g_1,m_1+2,n_1}] 
(\GG|\Phi_{NS}\rangle, |\Psi_{NS}\rangle^{\otimes m_1}, |\tilde\vp_r\rangle,
|\Psi_R\rangle^{\otimes n_1}) 
\nonumber \\ &&
\times \int_{\RR_{g_2,m_2+1,n_2}}
\Omega^{(g_2,m_2+1,n_2)}_{6g_2
-6+2 (m_2+n_2+1)}
(\GG|\tilde\vp^r\rangle, |\Psi_{NS}\rangle^{\otimes m_2}, |\Psi_R\rangle^{\otimes n_2}) 
\nonumber \\ &&
- \sum_{g_1,g_2} g_s{}^{2(g_1+g_2)} \, 
\sum_{m_1,m_2,n_1,n_2\ge 0}{1\over m_1! m_2! n_1! n_2!}  \nonumber \\ &&
\times \int_{\RR_{g_1,m_1+2,n_1}}
\Omega^{(g_1,m_1+2,n_1)}_{6g_1
-6+2 (m_1+n_1+2)}
(\GG|\Phi_{NS}\rangle, |\Psi_{NS}\rangle^{\otimes m_1}, |\tilde\vp_r\rangle,
|\Psi_R\rangle^{\otimes n_1}) 
\nonumber \\ &&
\times \int_{\RR_{g_2,m_2+1,n_2}}
\Omega^{(g_2,m_2+1,n_2)}_{6g_2
-5+2 (m_2+n_2+1)}[\wh U_{g_2,m_2+1,n_2}] 
(\GG|\tilde\vp^r\rangle, |\Psi_{NS}\rangle^{\otimes m_2}, |\Psi_R\rangle^{\otimes n_2}) 
%\nonumber \\ &&
\een
where we have used the fact that we get an extra minus sign from the interchange of the
operation of integrating over the angular variable $\theta$ of the plumbing 
fixture relation and
contraction with a vector field $\wh U$.
This cancels a minus sign coming from the $\prod_i \tilde\sigma_i$ factors in \refb{efactor}.
Note also that to maintain uniformity with the corresponding analysis involving terms with
Ramond degeneration, we have inserted a factor of $\GG$ in front of the basis states
$|\tilde\vp^r\rangle$ of $\HH_{NS}$ even though acting on NS sector states $\GG$ reduces
to the identity operator. Using \refb{edefcurly} and \refb{edefsquare}, 
eq.\refb{enname} can be rewritten as
\ben \label{ei4}
&& I_4 = -\sum_{g_1} g_s{}^{2g_1} \, 
\sum_{m_1,m_2,n_1,n_2\ge 0}{1\over m_1! m_2! n_1! n_2!}  \nonumber \\ &&
\times \int_{\RR_{g_1,m_1+2,n_1}}
\Omega^{(g_1,m_1+2,n_1)}_{6g_1
-5+2 (m_1+n_1+2)}
[\wh U_{g_1,m_1+2,n_1}] 
(\GG|\Phi_{NS}\rangle, |\Psi_{NS}\rangle^{\otimes m_1}, 
\GG[(\Psi_{NS})^{m_2} (\Psi_R)^{n_2}]_{NS}, |\Psi_R\rangle^{\otimes n_1})
\nonumber \\
&& -\sum_{g_2} g_s{}^{2g_2} \, 
\sum_{m_1,m_2,n_1,n_2\ge 0}{1\over m_1! m_2! n_1! n_2!}  \nonumber \\ &&
\times \int_{\RR_{g_2,m_2+1,n_2}}
\Omega^{(g_2,m_2+1,n_2)}_{6g_2
-5+2 (m_2+n_2+1)}[\wh U_{g_2,m_2+1,n_2}] 
(\GG[(\GG\Phi_{NS}) (\Psi_{NS})^{m_1} (\Psi_R)^{n_1}]_{NS}, 
|\Psi_{NS}\rangle^{\otimes m_2}, |\Psi_R\rangle^{\otimes n_2})\,. \nonumber \\
\een
We now see that the first term on the right hand side of \refb{ei4}, when added to $I_2$
defined in \refb{edeltaactionbb},
vanishes after using equations of motion. On the other hand the second term on the
right hand side of \refb{ei4} is equal to $J_2$ defined in \refb{ejj}. Thus we have
\be \label{eq12}
I_2 + I_4 - J_2=0\, .
\ee
In exactly the same way one can show that
\be \label{eq14}
I_3+I_5 - J_3=0\, .
\ee
Combining \refb{eq1}, \refb{eq12} and \refb{eq14} 
we see that upon using equation of motion we have
\be
(I_1+I_2+I_3+I_4+I_5) - (J_1+J_2+J_3)=0\, .
\ee
A similar analysis can be carried out involving the
terms involving $|\Phi_{R}\rangle$. This establishes \refb{edetde}.

Using the results above one can now show in a straightforward manner that the
renormalized mass is independent of the choice of the local coordinates and PCO
locations used to define the off-shell amplitudes and the 1PI effective theory. 
For this we use the fact that to determine the renormalized physical masses we need
to examine the zero eigenvalues of the gauge invariant kinetic operator giving the linearized
equations of motion around the classical solution $|\Psi_{cl}\rangle$ representing
the vacuum. The zero eigenvalues which exist for all values of the
momentum vector $k$ correspond to
pure gauge states. On the other hand 
the zero eigenvalues which exist only for special values of
$k^2$ correspond to physical states and the values of $-k^2$ at which the zero
eigenvalues appear give the physical renormalized mass$^2$. As discussed in detail
in \cite{1411.7478}, the fact that a change in local coordinates and 
PCO locations correspond to 
a field redefinition means that these transformations do not affect the (non-)existence 
of the zero eigenvalue at a given momentum, -- they only change the form of the
corresponding eigenstate. Thus the values of $-k^2$ at which the zero eigenvalues
appear remain unchanged, showing that the physical renormalized mass$^2$ are not
affected by the change in the choice of local coordinates and/or PCO locations.

\sectiono{Space-time supersymmetry} \label{ssusy}

In heterotic string theory
the gauge transformation parameter $|\Lambda\rangle$ appearing in
\refb{egauge} contains an R sector component $|\Lambda_R\rangle$ which is a state
of ghost number 1 and picture number $-1/2$. This includes local supersymmetry
transformations, just as $|\Lambda_{NS}\rangle$ includes
general coordinate and local
gauge transformations. In type II theories the local supersymmetry transformations
are contained in $|\Lambda_{RNS}\rangle$ and $|\Lambda_{NSR}\rangle$.

Our interest here is in understanding global supersymmetry transformations. These are
special choices of $|\Lambda\rangle$ which leave the vacuum invariant.  Our
goal will be to develop a systematic procedure for finding such $|\Lambda\rangle$'s.
For definiteness we focus on the heterotic string theory -- generalization to type II string
theories is straightforward. We begin by recalling that in general
the vacuum is not given by 
the $|\Psi\rangle=0$ configuration -- instead it corresponds to some specific 
configuration $|\Psi_{cl}\rangle$ whose systematic construction was described
in \cite{1411.7478}. 
Thus the global supersymmetry transformation parameter $|\Lambda_R\rangle$,
which by definition will be taken to carry zero momentum,
must satisfy
\be
Q_B|\Lambda_R\rangle + \sum_{n=0}^\infty {1\over n!} \XX_0 [\Psi_{cl}^n \Lambda_R]=0\, .
\ee 
We can look for a solution in a power series in the string coupling $g_s$ by beginning 
with the leading order solution
\be
|\Lambda_0\rangle = |c e^{-\phi/2}S_\alpha\rangle\, ,
\ee
where $S_\alpha$ is an appropriate dimension $(0,5/8)$ spin field from the matter sector. We
then compute the corrections iteratively by solving\footnote{Even though the usual
perturbation expansion in string theory is in powers of $g_s{}^2$, we take the expansion
to be in powers of $g_s$ since in some cases $|\Psi_{cl}\rangle$ may be of order 
$g_s$\cite{1404.6254,1411.7478}.}
\be \label{esusy1}
Q_B|\Lambda_{k}\rangle = - \sum_{n=0}^\infty {1\over n!} \XX_0 [\Psi_{cl}^n \Lambda_{k-1}]
+ \OO(g_s{}^{k+1})\, .
\ee
Here $|\Lambda_k\rangle$ is the approximation to the global supersymmetry transformation
parameter $|\Lambda_R\rangle$ to order $g_s{}^k$.
For consistency we need to ensure that if $|\Lambda_{k-1}\rangle$ satisfies 
the above equation with $k$ replaced by $(k-1)$, then we must have
\be \label{e6.4}
Q_B \sum_{n=0}^\infty {1\over n!} \XX_0 [\Psi_{cl}^n \Lambda_{k-1}] = \OO(g_s{}^{k+1})
\, .
\ee
The proof of this involves straightforward manipulation using \refb{emain} and
the equations of motion \refb{eeom} for $|\Psi_{cl}\rangle$ and
was given in \cite{1411.7478} in a slightly different context while discussing mass renormalization.
An obstruction arises if the right hand side of \refb{esusy1}, which is a state of
ghost number 2 and picture number $-1/2$, contains a non-trivial element of the
BRST cohomology; this is allowed by \refb{e6.4}.  
A basis of such states is provided by the zero momentum states of
physical massless fermions. If the right hand side of \refb{esusy1} has non-zero
component along such a state then supersymmetry is broken and the corresponding
state represents the zero momentum goldstino state associated with the
broken supersymmetry.
Thus the condition for unbroken supersymmetry will require that the 
right hand
side of \refb{esusy1} does not have any component along these possible goldstino
states. Since the BPZ inner product with $c_0^-$ insertion pairs states of ghost number 2
and picture number $-1/2$ with states of ghost number $3$ and picture number
$-3/2$, the above condition can also be expressed as
\be \label{esusy2}
\langle \phi | c_0^- \sum_{n=0}^\infty {1\over n!} \XX_0 |[\Psi_{cl}^n \Lambda_{k-1}]\rangle
=\OO(g_s{}^{k+1})\, ,
\ee
for any BRST invariant 
state $|\phi\rangle$ of ghost number 3 and picture number $-3/2$.
The possible non-trivial constraints come from states with $L_0^+$ eigenvalue 0, since
BRST invariant states with $L_0^+\ne 0$ are also BRST trivial.

We can make connection with the criteria given in \cite{1209.5461,1304.2832} 
by noting that if such obstructions
are absent up to a given order then to that order we can solve \refb{esusy1} by taking
\be \label{esolam}
|\Lambda_k\rangle = - \sum_{n=0}^\infty {1\over n!} {b_0^+\over L_0^+} (1 - {\bf P})
\XX_0 [\Psi_{cl}^n \Lambda_{k-1}] + |\tau_k\rangle\, ,
\ee
where ${\bf P}$ denotes the projection operator into $L_0^+=0$ states
and $|\tau_k\rangle$ is an $L_0^+=0$  state satisfying
\be
Q_B |\tau_k\rangle = - \sum_{n=0}^\infty {1\over n!} {\bf P}
\XX_0 [\Psi_{cl}^n \Lambda_{k-1}] \, .
\ee
In order to continue this construction to the next order 
we need that \refb{esusy2} should hold with $k$ replaced by
$k+1$. If we replace ${\bf P}$ by 0 (i.e.\ ignore the
special treatment of the $L_0^+=0$ states) and make a similar operation in the
construction of $|\Psi_{cl}\rangle$ described in \cite{1411.7478}, 
then this condition can be shown to
be equivalent to the requirement that the full two point function 
of $|\phi\rangle$ and
$|\Lambda_0\rangle$ -- including 1PI and 1PR contributions --
vanish to order $g_s{}^{k+1}$.  This is precisely the condition
for unbroken supersymmetry described in \cite{1209.5461,1304.2832}. 
However this two point function has
to be regularized to deal with divergences associated with separating type degenerations
of Riemann surfaces at the intermediate stage of the computation. In contrast our 
construction removes these divergences from the beginning by inserting the projection
operator $(1-{\bf P})$ in \refb{esolam} and compensates for this by inclusion of the
additional state
$|\tau_k\rangle$ (and similar operation in the construction of $|\Psi_{cl}\rangle$
described in \cite{1411.7478}.)

\sectiono{Discussion} \label{sdiss}

We conclude the paper by discussing possible applications of the formalism developed
here beyond studying the problems of mass renormalization and vacuum shift.

\begin{enumerate}
\item The requirement of gauge invariance puts strong constraint on the 1PI effective theory.
Indeed the identity \refb{eboundary} which is crucial for proving the infinite dimensional
gauge invariance of the 1PI effective
theory is also responsible for the fact that the different subspaces of $\wt\PP_{g,m,n}$
associated with different Feynman diagrams of this theory
fit together to give the full integration cycle.
Given this, one might wonder if the infinite dimensional gauge symmetry could also be
useful for constraining the non-perturbative corrections to the 1PI effective theory. It is
worth examining this question further since the current approach to the study of
non-perturbative effects in string theory is based mostly on the intuitions from the low
energy theory.
\item Formulating string theory in the RR background has been an open problem. The
1PI effective theory could provide a way out for weak RR background, since we could
construct the 1PI theory in a background where there is no RR field, and then
study the effect of switching on RR background by expanding the original 1PI equations
of motion around the new background in powers of the RR background
field. This could for example
give a way to study $g_s$ and $1/L$ corrections systematically in string theory on $AdS$ spaces
of size $L$, 
since
typically in the large $L$ limit the RR field strength is small (locally).
\end{enumerate}

\bigskip

{\bf Acknowledgement:}
We thank Roji Pius, Arnab Rudra, Edward Witten
and Barton Zwiebach for useful discussions and Barton Zwiebach for his
valuable comments on an earlier version of the manuscript.
This work  was
supported in part by the 
DAE project 12-R\&D-HRI-5.02-0303 and J. C. Bose fellowship of 
the Department of Science and Technology, India.

%\small

%\baselineskip 14pt

\end{document}